\documentclass[10pt]{article}

\usepackage[superscript,sort]{cite}

\usepackage{ifthen,ifpdf}

\ifpdf
	\usepackage[pdftex]{hyperref}	\hypersetup{colorlinks=true,linkcolor=black,citecolor=black,urlcolor=blue,backref=page,
	bookmarks=true, breaklinks=true,plainpages=false }
	\usepackage[pdftex]{graphicx}
	\usepackage[usenames,pdftex]{color}
\else
	\usepackage[colorlinks=true,breaklinks=true,plainpages=false]{hyperref}
	\usepackage{graphicx}
	\usepackage[usenames]{color}
\fi

\usepackage{amsthm,amscd,amsxtra,amsfonts,amsmath,amssymb,multirow, bm}
\usepackage{wrapfig}
\usepackage[footnotesize]{caption}
\usepackage[tiny,compact]{titlesec}
\usepackage{subcaption}

\usepackage[textwidth=0.8in,textsize=footnotesize]{todonotes}

\usepackage{helvet}

\usepackage{soul}

 \usepackage{colortbl}

\usepackage[utf8]{inputenc}

\usepackage[letterpaper,top=0.5in,bottom=0.5in,left=0.5in,right=0.5in]{geometry}
\usepackage{wrapfig}
\usepackage[tiny,compact]{titlesec}
\usepackage{xspace}
\usepackage{bm}
\usepackage{hyperref}
\usepackage{url,color}
\usepackage{multirow}
\usepackage{tabularx}
\usepackage{booktabs}
\usepackage{tikz}
\usepackage{curves}
\usepackage{caption}
\usepackage{placeins} 
\titlespacing*{\section}{0pt}{*0}{*0}
\titlespacing*{\subsection}{0pt}{*0}{*0}
\titlespacing*{\subsubsection}{0pt}{*0}{*0}
\titlespacing{\paragraph}{0pt}{*0}{*1}
\definecolor{MyPurple}{rgb}{1,0,1}

\DeclareMathOperator*{\argmin}{arg\,min}

\begin{document}




\title{A review of mathematical representations of biomolecular data
}

\author{Duc Duy Nguyen$^{1}$,  Zixuan Cang$^{1}$ and  Guo-Wei Wei$^{1,2,3,}$\footnote{
		Corresponding to Guo-Wei Wei.		Email: weig@msu.edu}\\
	$^1$ Department of Mathematics,
	Michigan State University, MI 48824, USA.\\
	$^2$ Department of Biochemistry and Molecular Biology,
	Michigan State University, MI 48824, USA. \\
$^3$ Department of Electrical and Computer Engineering,
	Michigan State University, MI 48824, USA. \\
}
\date{\today}
\maketitle

\begin{abstract}
Recently, machine learning (ML) has established itself in various worldwide benchmarking competitions in computational biology, including Critical Assessment of Structure Prediction (CASP) and Drug Design Data Resource (D3R) Grand Challenges. However, the intricate structural complexity and high ML dimensionality of biomolecular datasets obstruct the efficient application of ML algorithms in the field. In addition to data and algorithm, an efficient ML machinery for biomolecular predictions must include structural representation as an indispensable component. Mathematical representations that simplify the biomolecular structural complexity and reduce ML dimensionality have emerged as a prime winner in D3R Grand Challenges. This review is devoted to the recent advances in developing low-dimensional and scalable mathematical representations of biomolecules in our laboratory. We discuss three classes of mathematical approaches, including algebraic topology, differential geometry, and graph theory. We elucidate how the physical and biological challenges have guided the evolution and development of these mathematical apparatuses for massive and diverse biomolecular data. We focus the performance analysis on the protein-ligand binding predictions in this review although these methods have had tremendous success in many other applications, such as protein classification,  virtual screening,  and the predictions of solubility, solvation free energy, toxicity, partition coefficient, protein folding stability changes upon mutation, etc.

\end{abstract}
 Key words:
Machine learning, deep learning, descriptors, binding data, algebraic topology, differential geometry, graph theory 
 {\setcounter{tocdepth}{4} \tableofcontents}
 \pagebreak
\section{Introduction}

 Recently, Google’s DeepMind has caught the world’s breath in winning the 13$^{\rm th}$ Critical Assessment of Structure Prediction (CASP13) competition using its latest artificial intelligence (AI) system, AlphaFold \cite{alquraishi2019alphafold}. The goal of the CASP is to develop and recognize the state-of-the-art technology in constructing protein three-dimensional (3D) structure from protein sequences, which are abundantly available nowadays. While many people were surprised by the power of AI when AlphaGo beat humans for the first time in the highly intelligent Go game a few years ago, it was not clear whether AI could tackle scientific challenges. Since CASP has been regarded as one of the most important challenges in computational biophysics, AlphaFold’s dominant win of 25 out of 43 contests ushers a new era of scientific discovery.

The algorithms underpinning ALphaFold's AI system are machine learning  (ML), including deep learning (DL). Indeed, ML is one of the most transformative technologies in history.  The combination of big data and ML has been referred to as both the “fourth industrial revolution” \cite{schwab2017fourth} and the “fourth paradigm of science” \cite{agrawal2016perspective}. However, this two-element combination may not work very well for biological science, particularly, biomolecular systems because of the intricate structural complexity and the intrinsic high dimensionality of biomolecular datasets \cite{butler2018machine}.  For example, a typical human protein-drug complex has so many possible configurations that even if a computer enumerates one possible configuration per second, it would still take longer than the universe has existed to reach the right configuration. The chemical and pharmacological spaces of drugs are so large that even all the world's computers put together do not have enough power for automated de novo drug design due to additional requirements in solubility, partition coefficient, permeability, clearance, toxicity, pharmacokinetics, and pharmacodynamics,  etc.

\begin{wrapfigure}{c}{1.50in}
\vspace{-4mm}
\includegraphics[keepaspectratio,width=1.50in]{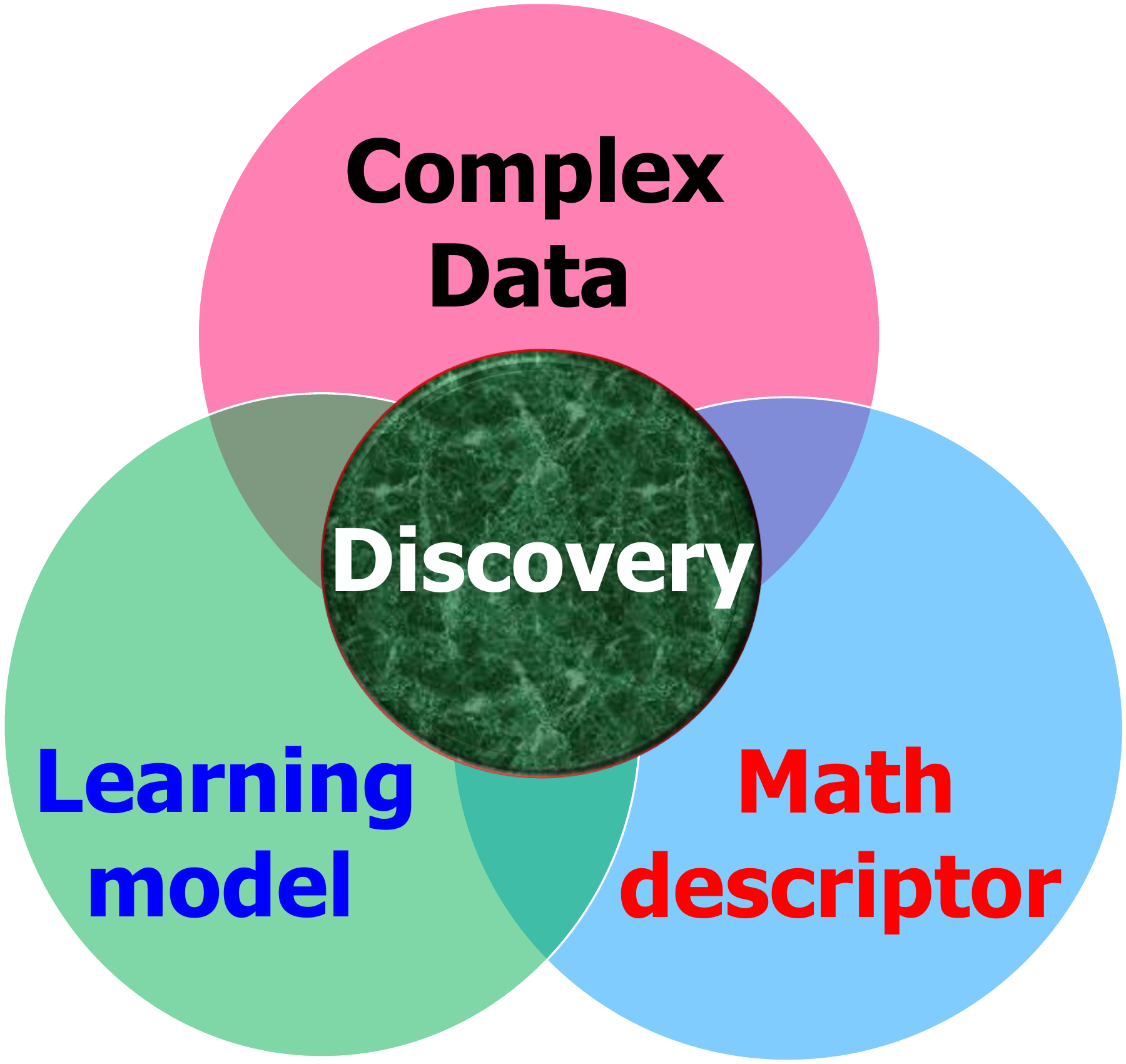}
\caption{ Illustration of essential elements for machine learning (ML) based discovery from complex biomolecular data.
      }
	\label{fig:learning}		
	\vspace{3mm}
\end{wrapfigure}
\begin{wrapfigure}{r}{3.0in}
  \vspace{-7mm}
\begin{center}
\includegraphics[keepaspectratio,width=3.in]{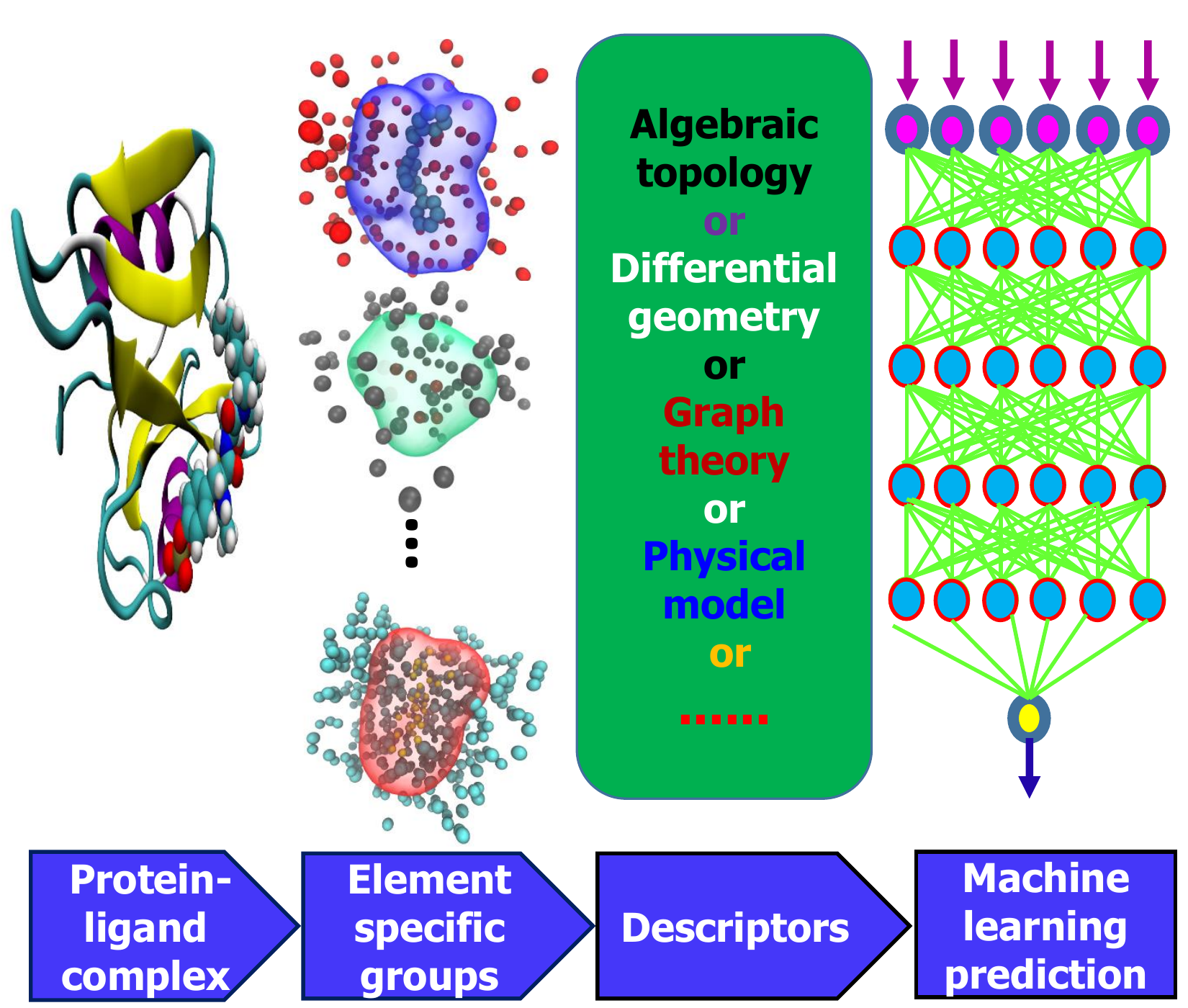}
\end{center}
 \vspace{-2mm}
\caption{Illustration of descriptor-based learning processes.}
\label{fig:Learningprocess}
 \vspace{-2mm}
\end{wrapfigure}

An appropriate low-dimensional representation of biomolecular structures  is required \cite{butler2018machine,brandt2018machine,Darnell:2008,huang2016communication,winter2019learning, wei2019system }  to translate the complex structural information into machine learning feature vectors or mathematical descriptors as shown in Fig. \ref{fig:Learningprocess}. As a result, various machine learning algorithms, particularly relatively simple ones without complex internal structures, can work efficiently and robustly with biomolecular data. 

   Descriptors or fingerprints are indispensable even for small molecules -- they play a fundamental role in quantitative structure-activity relationship (QSAR) and quantitative structure-property relationships (QSPR)  analysis, virtual screening, similarity-based compound search, target molecule ranking, drug absorption, distribution, metabolism, and excretion (ADME) prediction, and other drug discovery processes. Molecular descriptors are property profiles of a molecule, usually in the form of vectors with each vector component indicating the existence, the degree or the frequency of one certain structure feature \cite{geppert2010current,roy2012electrotopological,tareq2010predictions}. Various descriptors have been developed in the past few decades \cite{rogers2010extended,lo2018machine,cereto2015molecular}. Most of them are 2D ones that can be extracted from molecular simplified molecular-input line-entry system (SMILES) strings without 3D structure information.  High dimensional descriptors have also been developed to utilize  3D molecular structures and other chemical and physical information \cite{verma20103d}. There are four main categories of 2D descriptors: 1) substructure keys-based fingerprints, 2) topological or path-based fingerprints, 3) circular fingerprints, and 4) pharmacophore fingerprints. Substructure keys-based fingerprints, such as molecular access system (MACCS) \cite{durant2002reoptimization},  are bit strings representing the presence of certain substructures or fragments from a given list of structural keys in a molecule. 
Topological or path-based descriptors, e.g., FP2 \cite{o2011open}, Daylight \cite{daylight} and electro-topological state (Estate) \cite{hall1995electrotopological},  are designed to analyze all the fragments of a molecule following a (usually linear) path up to a certain number of bonds, and then hashing every one of these paths to create fingerprints.
Circular fingerprints, such as extended-connectivity fingerprint (ECFP) \cite{rogers2010extended},
 are also hashed topological fingerprints but rather than looking for paths in a molecule, they record the environment of each atom up to a pre-defined radius. 
Pharmacophore fingerprints include the relevant features and interactions needed for a molecule to be active against a given target, including  2D-pharmacophore \cite{landrum2006rdkit}, 3D-pharmacophore \cite{landrum2006rdkit} and extended reduced graph (ERG) \cite{stiefl2006erg} fingerprints as examples.

 However, typically designed for 2D SMILES strings, the aforementioned small-molecular descriptors do not work well for macromolecules that have   complex 3D structures.  The complexity of biomolecular structure, function, and dynamics often makes the structural representation inclusive, inadequate,  inefficient and sometimes intractable.  These challenges call for innovative design strategies for the representation of macromolecules.

Popular molecular mechanics models use bonded terms for describing covalent bond interactions and non-bonded terms for representing long-range electrostatic and van der Waals effects.  As a result, the early effort has been focused on exploring related {\it physical descriptors} to account for hydrogen bonds, electrostatic effects, van der Waals interactions, hydrophilicity, and hydrophobicity. These descriptors have been applied to many macromolecular systems, such as protein-protein interaction hot spots \cite{Darnell:2008,Demerdash:2009,huang2016communication,lu2019predicting}.  Similar physical descriptors in terms of van der Waals interaction, Coulomb interaction, electrostatic potential,  electrostatic binding free energy,  reaction field energy,  surface areas, volumes, etc, were applied by us to predictions of protein-ligand affinity \cite{BaoWang:2017FFTB } and solvation free energy \cite{BaoWang:2016HPK, BaoWang:2018FFTS }.
 However, the major limitation of physical descriptors is that they highly depend on existing molecular force fields, such as  radii, charges,   polarizability, dielectric constant, and van der Waals well depth, and thus could inherit errors from upstream physical models. As a result, these descriptors are often not as competitive as state-of-art force-field-free   models based on advanced mathematics \cite{wei2019system,nguyen2019mathematical}.

\begin{wrapfigure}{c}{4.2in}
\vspace{-2mm}
\includegraphics[keepaspectratio,width=4.2in]{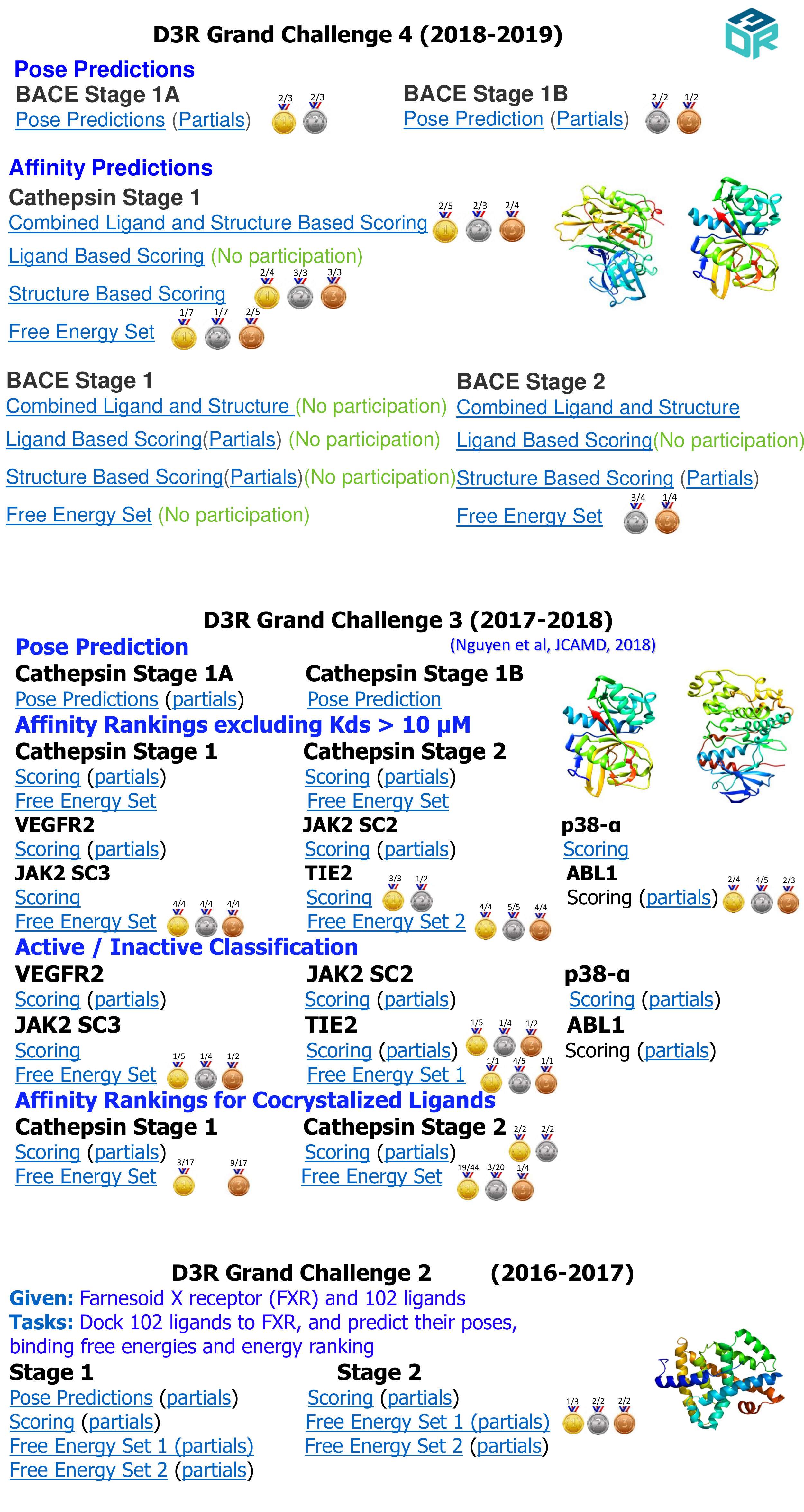}
\caption{Wei team's performance in D3R Grand Challenges 2, 3 and 4 \cite{nguyen2019mathematical,nguyen2019mathdl}, community-wide competition series in computer-aided drug design, with components addressing blind predictions of pose-prediction, affinity ranking, and binding free energy. The golden medal, silver medal, and bronze medal label the contest where our prediction was ranked 1$^{\rm st}$, 2$^{\rm nd}$, and 3$^{\rm rd}$, respectively.
      }
	\label{fig:D3RGC2}
	\vspace{-3mm}
\end{wrapfigure}

Topology analyzes space, connectivity, dimension, and transformation. Topology offers the highest level of abstraction and thus could provide an efficient tool for tackling high-dimensional biological data \cite{Schlick:1992trefoil,Zomorodian:2005,sumners:1992}. However, topology typically oversimplifies geometric information.
Persistent homology is a new branch of algebraic topology that is able to bridge geometry and topology  \cite{edelsbrunner2000topological,Zomorodian:2005,Zomorodian:2008}.  This approach has been applied to macromolecular analysis \cite{YaoY:2009,Gameiro:2014,KLXia:2014c,KLXia:2015a,KLXia:2015b,KLXia:2015c,KLXia:2015d,KLXia:2015e, ZXCang:2015, BaoWang:2016a,ESES:2017}.
Nonetheless, it neglects critical chemical/biological information when it is directly applied to complex biomolecular structures. Recently, we have introduced element-specific persistent homology to retain critical  biological information during the topological abstraction, rendering a potentially  revolutionary descriptor for  biomolecular data \cite{ ZXCang:2017a,ZXCang:2017b,ZXCang:2017c,ZXCang:2018a}.

 Graph theory studies the modeling of pairwise relations between vertices or nodes \cite{FanR.K.Chung1997b}. 
Geometric graphs admit geometric objects as graph nodes while algebraic graphs utilize algebraic techniques to study the relations between nodes. Both geometric graph theory and algebraic graph theory have been widely applied to biomolecular systems \cite{Twarock:2008,janezic2015graph,li2018machine,winter2019learning}. For example, spectral graph theory has been used to represent protein C$_\alpha$ atoms as an elastic mass-and-spring network in Gaussian network model (GNM) \cite{Bahar:1997} and anisotropic network model (ANM) \cite{Atilgan:2001}. Extremal graph theory concerns unavoidable patterns and structures in graphs with given density or distribution. It has potential applications to   chromosome packing and Hi-C data. 
However, most graph theory methods suffer from the neglecting of critical biological information and non-covalent interactions, and sometimes,  inappropriate distance metrics for biomolecular interactions.
In the past few year, we have developed weighted graphs \cite{KLXia:2013d,  KLXia:2014b,KLXia:2017, Opron:2014,Opron:2015a,Opron:2016a, DDNguyen:2016b}, multiscale graphs \cite{Opron:2015a,KLXia:2015f}, and  colored graphs  \cite{DDNguyen:2017d,DBramer:2018a} for modeling biomolecular systems.   These new graph theory methods are found to be some of the most powerful descriptors of macromolecules  \cite{DDNguyen:2017d,DBramer:2018a,nguyen2019agl}.

How biomolecules assume complex structures and intricate shapes and why biomolecular complexes admit convoluted interfaces between different parts can be naturally described by differential geometry, a mathematical subject drawing on differential calculus, integral calculus,  algebra,  and differential equation to study problems in geometry or differentiable manifolds. Einstein used this approach to formulate his general theory of relativity. Curve and curvature analysis has been applied to the shape analysis of molecular surfaces \cite{duncan1993shape} and protein folding trajectories \cite{SFOW08,CW13}. In the past two decades, we have developed a variety of  differential geometry   models for biomolecular surface analysis  \cite{Wei:2005,Bates:2006,Bates:2006f,Bates:2008,XFeng:2012a,XFeng:2013b,KLXia:2014a},  solvation  modeling  \cite{ZhanChen:2010a, ZhanChen:2010b, ZhanChen:2011a, ZhanChen:2012, DuanChen:2012a, DuanChen:2012b, Wei:2012, Daily:2013, Thomas:2013,DDNguyen:2016c}, ion-channel study \cite{Wei:2009,Wei:2012, Wei:2013,  DuanChen:2012a, DuanChen:2012b}, protein binding pocket detection \cite{RDZhao:2018a}, and protein-ligand binding affinity prediction \cite{nguyen2019dg}.  Differential geometry-based descriptors are able to offer a high-level abstraction of macromolecular structures \cite{nguyen2019dg}.

We have pursued  differential geometry, algebraic topology, graph theory and other mathematical methods, such as de Rham-Hodge theory \cite{zhao20193d,zhao2019rham}, for modeling, analysis and characterization of biomolecular systems for near two decades. Using these methods, we have studied a number of biomolecular systems and problems, including macromolecular electrostatics, implicit solvent models, ion channels, protein flexibility, geometric analysis, surface modeling, and multiscale analysis.
Our formulations have evolved and improved over time. In 2015, we proposed one of the first integration of persistent homology and machine learning and applied this new approach to protein classification. Since then, we have demonstrated  the superiority of our mathematical descriptors over other existing methods in a wide variety of other applications, including the predictions of protein thermal fluctuations \cite{Opron:2014,Opron:2015a, KLXia:2015f, DBramer:2018a},   toxicity \cite{KDWu:2018a}, protein-ligand binding affinity \cite{ZXCang:2017b, DDNguyen:2017d,BaoWang:2017FFTB, nguyen2019dg,nguyen2019agl}, mutation-induced protein stability changes \cite{ZXCang:2017a,ZXCang:2017c},  solvation \cite{ZhanChen:2012, BaoWang:2015a, BaoWang:2016HPK,BaoWang:2018FFTS}, solubility \cite{KDWu:2018b}, partition coefficient \cite{KDWu:2018b} and virtual screening \cite{ZXCang:2018a}. As shown in  Fig. \ref{fig:D3RGC2},  the aforementioned mathematical approaches have enabled us to win many contests in   \href{https://drugdesigndata.org/about/grand-challenge }{D3R Grand Challenges}, a worldwide competition series in computer-aided drug design \cite{nguyen2019mathematical}.

 Due to the abstract nature of mathematical approaches and the fact that our results are scattered over a large number of subjects and topics it is difficult for the researcher who has no formal training in mathematics to use these methods. Therefore, there is a pressing need to elucidate these methods in physical terms, provide simplified descriptions,  and interpret their working principles. To this end, we provide a review of our mathematical descriptors.  Our goal is to offer a coherent description of these methods for protein-ligand binding interactions so that the reader can better understand how to use advanced mathematics for describing macromolecules and their interaction complexes.

Like small molecular descriptors, macromolecular descriptors, once designed, can be applied to different tasks in principle.  However, many different types of applications require specially designed macromolecular descriptors. For example, in protein B-factor prediction, one deals with the atomic property, while in predicting protein stability changes upon mutation, solubility, etc. one considers molecular properties. Additionally, in protein-ligand binding affinity predictions, one deals with the property of protein-ligand complexes.  Therefore, different mathematical descriptors are required to tackle atomic, molecular, and molecular complex properties.  Another complication is due to different systems. For example,  descriptors for the binding affinity of protein-ligand interactions should differ from those for the binding affinity of protein-protein interactions or protein-nucleic acid interactions. The other hindrance  arises from specific tasks. For example, protein classification, one concerns secondary structures and needs to design macromolecular descriptors to capture secondary structural differences.  In general, macromolecules and their interactive complexes are inherent of multiscale, multiphysics, multi-dynamics and multifunction. Their descriptions can vary from cases to cases. We cannot cover all possible situations in this review.

Biologically, protein-ligand binding interactions are tremendously important for living organisms. ligand-receptor agonist binding is known to initiate a vast variety of molecular and/or cellular processes, from transmitter-mediated signal transduction, hormone or growth factor regulated metabolic pathways, stimulus-initiated gene expression, enzyme production, to cell secretion. Therefore,  the understanding of protein-ligand binding interactions is a central issue in biological sciences, including drug design and discovery. Despite much research in the past, the molecular mechanism of protein-ligand binding interactions is still elusive. A prevalent view is that protein-ligand binding is initiated through protein-ligand molecular recognition, synergistic corporation, and conformational changes. Computationally, the prediction of protein-ligand binding affinity is sufficiently challenging.  Consequently, we focus on mathematical descriptors for protein-ligand binding affinity predictions to illustrate their design and application in the present review.

\section{Methods}\label{Methods}
In this section, we briefly review three classes of mathematical descriptors, i.e., descriptors constructed from algebraic topology, graph theory, and differential geometry.

\subsection{Algebraic topology-based methods}\label{Method:topology}

\subsubsection{Background}\label{sec:SimplicialHomology}

\begin{figure}
\begin{center}
\includegraphics[width=0.8\textwidth]{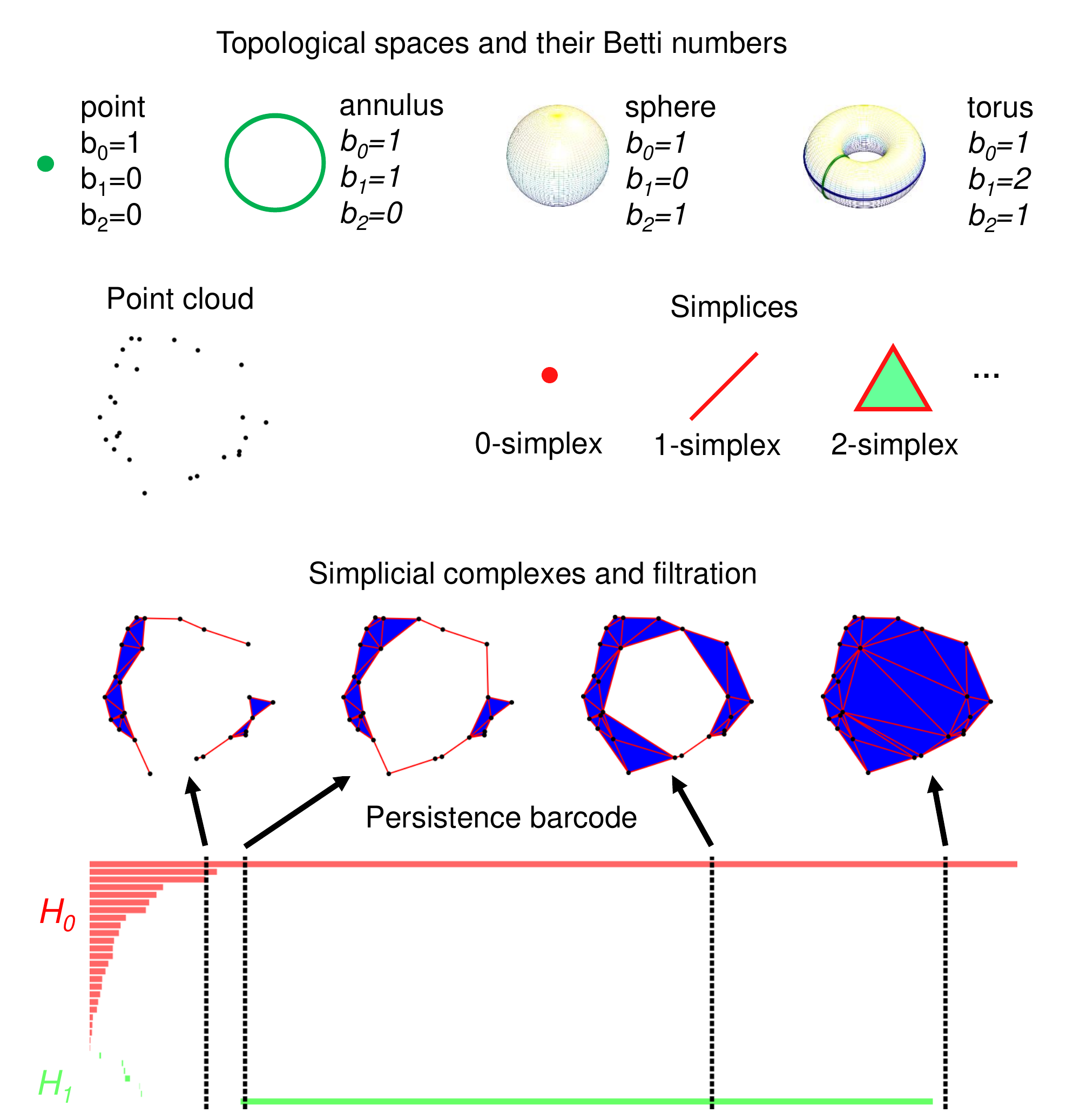}
\end{center}
\caption{Topological description of point clouds via persistent homology.}\label{fig:ph_basic}
\end{figure}

  Topology   dramatically simplifies geometric complexity \cite{Schlick:1992trefoil,Zomorodian:2005,sumners:1992,IKDarcy:2013,CHeitsch:2014,Demerdash:2009,DasGupta2016,XShi:2011}.
The study of topology deals with the connectivity of different components in space and characterizes independent entities,  rings, and higher dimensional faces within the space \cite{Kaczynski:2004}.  For example, simplicial homology, a type of algebraic topology, concerns the identification of topological invariants from a set of discrete node coordinates such as atomic coordinates in a protein. For a given (protein) configuration, independent components, rings, and cavities are topological invariants and their numbers are called Betti-0, Betti-1, and Betti-2, respectively, see Fig. \ref{fig:ph_basic}. To study topological invariants in a discrete dataset, simplicial complexes are constructed by gluing simplices under various settings, such as the Vietoris-Rips (VR) complex,  $\check{C}$ech complex or alpha complex.  Specifically, a 0-simplex is a vertex, a 1-simplex an edge, a 2-simplex  a triangle, and a 3-simplex  a tetrahedron, as illustrated in  Fig. \ref{fig:ph_basic}. Algebraic groups built on these simplicial complexes are used in simplicial homology to systematically compute various Betti numbers. There is also cubical complex \cite{Kaczynski:2004} built upon volumetric data, including those from  biomolecules \cite{BaoWang:2016a}.

\begin{figure}
\begin{center}
\includegraphics[width=1.0\textwidth]{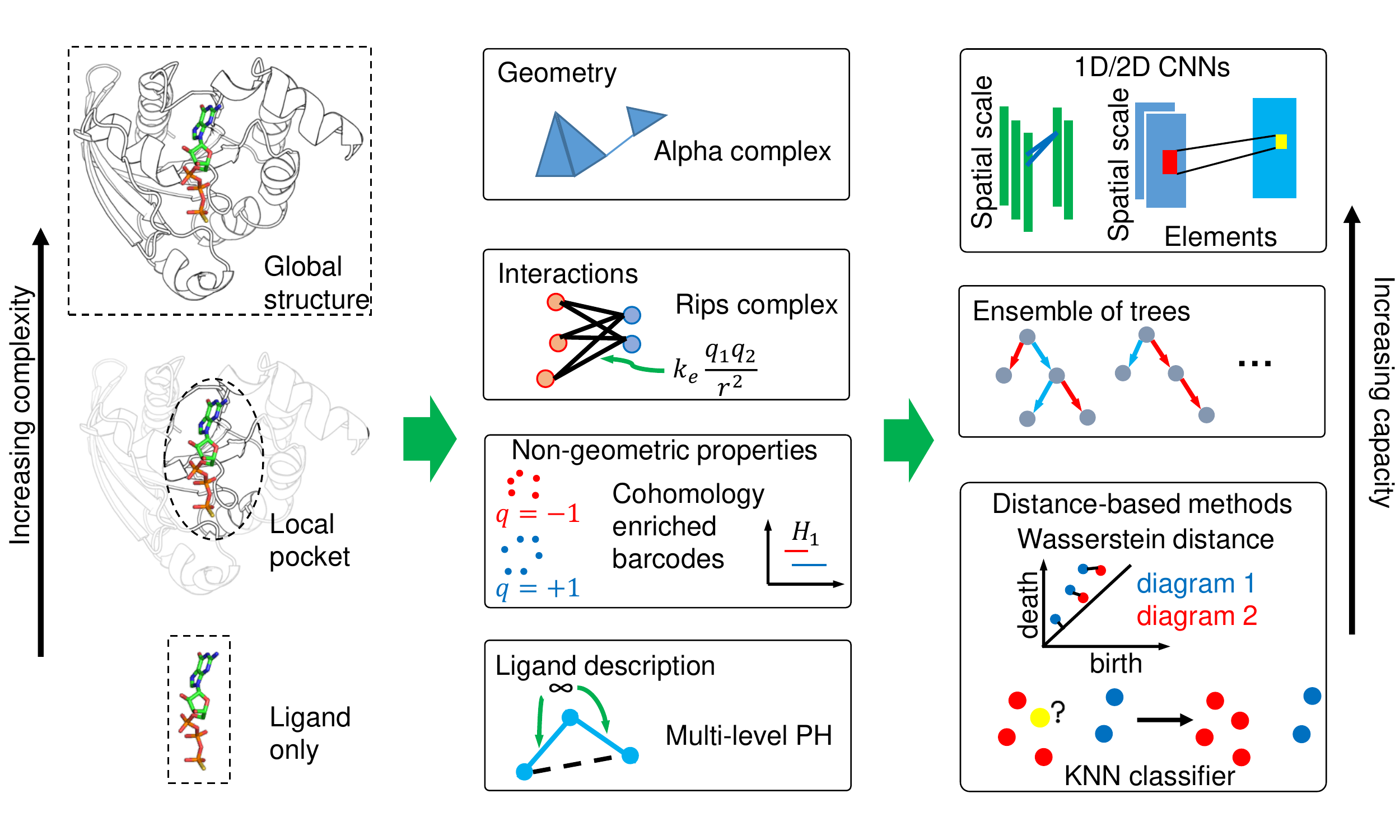}
\end{center}
\caption{Workflow of topology based protein-ligand binding affinity prediction.
In multi-level persistent homology, the distance between covalent bonds are set to $\infty$ to avoid their disturbance to the topological representation of non-covalent bonds.
}\label{fig:ph_workflow}
\end{figure}

However, conventional topology or homology is truly free of metrics or coordinates, and thus retains too little geometric information to be practically useful. Persistent homology is a relatively new branch of algebraic topology that embeds multiscale geometric information into topological invariants to achieve a topological description of geometric details \cite{edelsbrunner2000topological,Zomorodian:2005}.
 It creates a sequence of topological spaces of a given object by varying a  filtration parameter, such as the radius of a ball or the level set of a surface function as shown in  Fig. \ref{fig:ph_basic}.
As a result, persistent homology can capture topological structures continuously over a range of spatial scales. Unlike commonly used computational homology which results in truly metric free representations, persistent homology embeds essential geometric information in topological invariants, e.g., topological descriptors or barcodes \cite{CZOG05} shown in Fig.  \ref{fig:ph_basic},  so that ``birth"  and ``death" of isolated components, circles, rings,  voids or cavities can be monitored at all geometric scales by topological measurements. A schematic illustration of our persistent homology-based machine learning predictions is given in Fig.  \ref{fig:ph_workflow}.   Key concepts are briefly discussed below. More mathematical detail can be found in the literature \cite{Zomorodian:2005}, including ours \cite{KLXia:2014c, KLXia:2015a}.

\paragraph{Simplicial complex}
A simplicial complex is a topological space consisting of vertices (points), edges (line segments), triangles, and their high dimensional counterparts.  Based on the simplicial complex, simplicial homology can be defined and used to analyze topological invariants.
The essential building blocks of geometry induced simplicial complex are simplices. Specifically, let $v_0,v_1,v_2,\cdots,v_k$ be $k+1$ affinely independent points; a (geometric) $k$-simplex $\sigma^k=\{v_0,v_1,v_2,\cdots,v_k\}$ is the convex hull of these points in $\mathbb{R}^N$ ($N\geq k$), and can be expressed as
$$
\sigma^k=\left\{\lambda_0 v_0+\lambda_1 v_1+ \cdots +\lambda_k v_k \mid \sum^{k}_{i=0}\lambda_i=1;0\leq \lambda_i \leq 1,i=0,1, \cdots,k \right\}.
$$
An $i$-dimensional face of $\sigma^k$ is defined as the convex hull formed by the subset of $i+1$ vertices from $\sigma^k$ ($k\geq i$). 
Geometrically, a $0, 1, 2, $ and $3$-simplex corresponds to a vertex, an edge, a triangle, and a tetrahedron, respectively.
A simplicial complex $K$ is a finite set of simplices such that
any face of a simplex from  $K$  is also in  $K$ and  
the intersection of any two simplices in  $K$ is either empty or a face of both. 
The underlying space $|K|$ is a union of all the simplices of $K$, i.e.,  $|K|=\cup_{\sigma\in K} \sigma$.

\paragraph{Homology}
The basic algebraic structure, chain groups, are defined for simplicial complexes so that homology can be characterized. A $k$-chain $[\sigma^k]$ is a formal sum $\sum_{i}\alpha_i\sigma^k_i$ of $k$-simplices $\sigma^k_i$. 
The coefficients $\alpha_i$ are often chosen in an algebraic field (typically, $\mathbb{Z}_2$). 
The set of all $k$-chains of the simplicial complex $K$ together with addition operation forms an abelian group $C_k(K, \mathbb{Z}_2)$. The homology of a topological space is represented also by a series of abelian groups, constructed based on these spaces of chains connected by boundary operators.
The boundary operator on chains $\partial_k: C_k \rightarrow C_{k-1}$ are defined by linear extension from the boundary operators on simplices. Omitting orientation (a simplification by using $\mathbb{Z}_2$) the boundary of a $k$-simplex $\sigma^k=\{v_0,v_1,v_2,\cdots,v_k\}$ is
$\partial_k \sigma^k = \sum^{k}_{i=0} \{ v_0, v_1, v_2, \cdots, \hat{v_i}, \cdots, v_k \},
$
where $\{v_0, v_1, v_2, \cdots ,\hat{v_i}, \cdots, v_k\}$ is    the $(k-1)$-simplex excluding $v_i$ from the vertex set. 
A key property of the boundary operator is that $\partial_{k-1}\partial_k= \emptyset$ and $\partial_0= \emptyset$. 
The $k$-cycle group $Z_k$ and the $k$-boundary group $B_k$ are the subgroups of $C_k$ defined as,
$Z_k={\rm Ker}~ \partial_k=\{c\in C_k \mid \partial_k c=\emptyset\},~ 
 { B_k={\rm Im} ~\partial_{k+1}= \{ \partial_{k+1} c \mid c\in C_{k+1} \}.}
$

An element in the $k$-th cycle group $Z_k$ (or the $k$-th boundary group $B_k$) is called a $k$-cycle (or the $k$-boundary, resp.).  As the boundary of a boundary is always empty $\partial_{k-1}\partial_k= \emptyset$,  one has $B_k\subseteq Z_k \subseteq C_k$. Topologically, a $k$-cycle is a union of $k$ dimensional loops (or closed membranes).
The $k$-th homology group $H_k$ is the quotient group generated by the $k$-cycle group $Z_k$ and $k$-boundary group $B_k$: $H_k=Z_k/B_k$. Two $k$-cycles are called homologous if they differ by a $k$-boundary element. From the fundamental theorem of finitely generated abelian groups, the $k$-th homology group $H_k$ can be expressed as a direct sum,
$H_k= {Z}\oplus \cdots \oplus {Z} \oplus {Z}_{p_1}\oplus \cdots \oplus {Z}_{p_n}= {Z}^{\beta_k} \oplus {Z}_{p_1}\oplus \cdots \oplus {Z}_{p_n},
$
where $\beta_k$, the rank of the free subgroup, is the $k$-th Betti number. 
Here  $ {Z}_{p_i}$ is torsion subgroup with torsion coefficients $\{p_i| i=1,2,...,n\}$, powers of prime numbers. 
 The Betti number can be simply calculated by
$\beta_k = {\rm rank} ~H_k= {\rm rank }~ Z_k - {\rm rank}~ B_k.
$
The geometric interpretations of  Betti numbers in $\mathbb{R}^3$ are as follows: $\beta_0$ represents the number of isolated components, $\beta_1$ is the number of independent one-dimensional loops (or circles), and $\beta_2$ describes the number of independent two-dimensional voids (or cavities). Together, the Betti numbers { $\{\beta_0,\beta_1,\beta_2,\cdots \}$} describes the intrinsic topological property of a system.

\paragraph{Persistent homology}
For a simplicial complex $K$, a filtration is defined as a nested sequence of subcomplexes,
$\varnothing = K^0 \subseteq K^1 \subseteq \cdots \subseteq K^m=K.
$
Generally speaking, abstract simplicial complexes generated from a filtration give a multiscale topological representation of the original space, from which related homology groups can be evaluated to reveal topological features. Specifically, upon passing the previous sequence to homology, we obtain a sequence of vector spaces connected by homomorphisms:
$H_*(K^0) \to H_*(K^1) \to \cdots \to H_*(K^m)$.
  Following this sequence of homology groups, sometimes new homology classes are created (i.e., without pre-image under the map $H_*(K^i) \to H_*(K^{i+1})$), and sometimes certain homology classes are destroyed (i.e., they have trivial image under $H_*(K^j) \to H_*(K^{j+1})$). The concept of persistence is introduced to measure the ``life-time'' of such homological features. The results can be summarized in the \emph{persistence barcodes} (or equivalently \emph{persistence diagrams}), consisting of a set of intervals $[x,y)$ with the beginning and ending values representing the birth and death of homology classes.
The introduction of filtration is of essential importance and directly leads to the invention of persistent homology. Generally speaking,    abstract simplicial complexes generated from a filtration give a multiscale representation of the corresponding topological space, from which related homology groups can be evaluated to reveal topological features. Furthermore, the concept of persistence is introduced for long-lasting topological features. However, we have shown that short-lived topological features are also important for biomolecular systems \cite{KLXia:2014c}. The $p$-persistent of $k$-th homology group, $K^i$, is
\begin{eqnarray}
H^{i,p}_k=Z^i_k/(B_k^{i+p}\bigcap Z^i_k).
\end{eqnarray}
Through the study of the persistence pattern of these topological features, the so-called persistent homology is capable of capturing the intrinsic properties of the underlying space solely from a discrete point set.

\begin{figure}
\begin{center}
\includegraphics[width=0.8\textwidth]{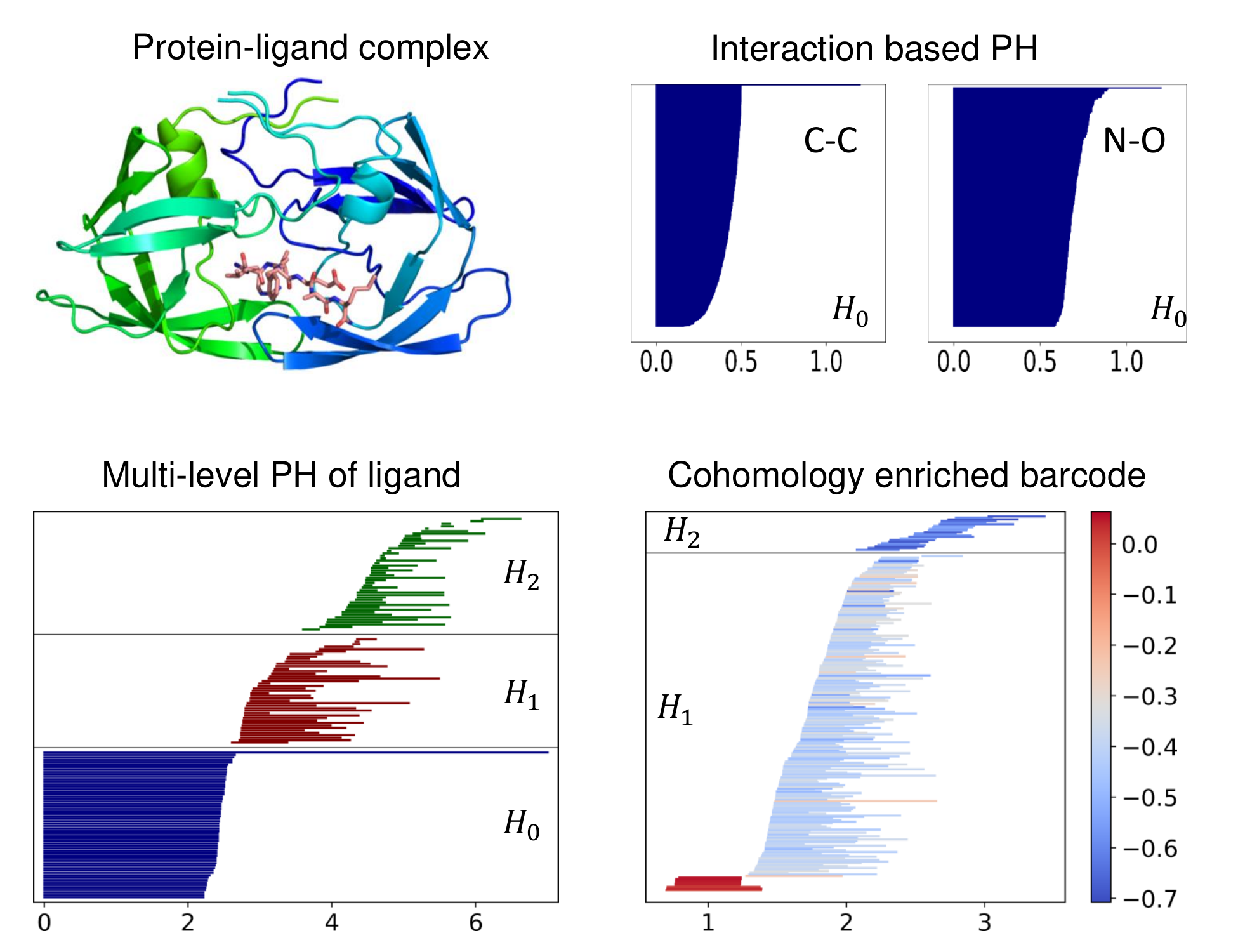}
\end{center}
\caption{Topological fingerprints addressing different aspects of the protein-ligand complex. (a) The example protein-ligand complex (PDB:1A94). (b) The $H_0$ barcodes from Rips filtration based on the Coulomb potential for carbon-carbon and nitrogen-oxygen interactions between protein and ligand. (c) The multi-level persistent homology characterization of the ligand revealing the non-covalent intramolecular interaction network. (d) The enriched barcode via persistent cohomology for atomic partial charges as the non-geometric information.}
\label{fig:ph_example}
\end{figure}

\paragraph{Filtration}
Given a set of discrete sample points, there are different ways to construct simplicial complexes. Typical constructions are based on the intersection patterns of the set of expanding balls centered at the sample points, such as \v{C}ech complex, (Vietoris-)Rips complex and alpha complex \cite{javaPlex,Mischaikow:2013}. The corresponding topological invariants, e.g., the Betti numbers, could be different depending on the choice of simplicial complexes. A common filtration for a set of atomistic data of a macromolecule is constructed by enlarging a common atomic radius $r$ from $0$. As the value of $r$ increases, the solid balls will grow and new simplices can be defined through the overlaps among the set of balls.
In Figure \ref{fig:ph_basic}, we illustrate this process by a set of points.
In Fig. \ref{fig:ph_example}, we demonstrate the persistent homology analysis of different aspects of a protein-ligand complex using the barcode representation.

\subsubsection{Challenge}

 Conventional topology and homology are independent of metrics or coordinates and thus retain too little geometric information to be practically useful in most biomolecular systems. While persistent homology incorporates more geometric information, it typically treats all atoms in a  macromolecule indifferently, which fails to recognize detailed chemical, physical,  and biological information  \cite{YaoY:2009,Gameiro:2014}. We introduced persistent homology as a {\it quantitative} tool for analyzing biomolecular systems  \cite{KLXia:2014c,KLXia:2015a,KLXia:2015b,KLXia:2015c,KLXia:2015d,KLXia:2015e,BaoWang:2016a,ESES:2017}. In particular, we introduced one the first topology-based machine learning algorithms for protein classification in 2015 \cite{ZXCang:2015}.   We further introduced element specific persistent homology, i.e., element-induced topology, to deal with massive and diverse bimolecular datasets  \cite{ZXCang:2015,ESES:2017,ZXCang:2017a,ZXCang:2017b,ZXCang:2017c}. Moreover, we introduced multi-level persistent homology to extract non-covalent-bond interactions \cite{ZXCang:2018a}. Furthermore, physics-embedded persistent homology was proposed to incorporate physical laws into topological invariants \cite{ZXCang:2018a}.
These new topological tools are potentially revolutionary for complex biomolecular data analysis \cite{wei2019system}.

\subsubsection{Element specific persistent homology}

Many types of interactions exist in a protein-ligand complex, for example, hydrophobic effects, hydrogen bonds, and electrostatics. Due to the mechanisms of these interactions, they happen under different geometric distances. Persistent homology, when applied to all the atoms, however, will mostly capture the interactions among nearest neighbors and hinder the detection of long-range interactions. Additionally,
it does not distinguish the difference between different element types and their combinations and thus, neglects important chemistry and biology.  Element specific persistent homology provides a simple yet effective solution to these issues. Instead of computing persistent homology for the whole molecule once, we perform persistent homology computations on a collection of subsets of atoms. For example, persistent homology on only carbon atoms characterizes the hydrophobic interaction network and the hydrogen bond interactions can be described by persistent homology on the set of nitrogen and oxygen atoms. Although different types of interactions have different characteristics, they may also influence each other. This encourages the iteration over all combinations of atom types which may result in large computation cost. Fortunately, as   Vietoris-Rips filtration is often used to characterize the interaction networks, we only need to generate the filtered simplicial complex once for all atoms and perform persistent homology computation on the subcomplexes of the filtered simplicial complex.

\subsubsection{Multi-level persistent homology}
Vietoris-Rips complex based only on pairwise distance is a widely used realization of filtration. When directly feeding the Euclidean distance between atoms to Rips complex construction, the interactions of interest such as electrostatic interactions can be flushed away by covalent bonds which usually have shorter lengths. This motivates us to incorporate a simple yet effective strategy to recover these important interactions by masking the original Euclidean distance matrix.  Specifically, we keep only the entries corresponding to the interaction of interest and set every other entry to infinity in the distance matrix. For example, we set distances between atoms from the same component (protein or ligand) to infinity to focus on the interactions between the protein and ligand. This strategy was found especially useful when dealing with ligands alone which often have a much simpler structure than the proteins or the protein-ligand complexes. We call this approach to small molecules \emph{multi-level persistent homology} of level $n$ where we set the distance between two atoms to infinity if the shortest path between them through the covalent bond network is at most of the length $n$. This treatment has led to powerful predictive tools in tasks only explicitly involving small molecules \cite{ZXCang:2018a,KDWu:2018a}.

\subsubsection{Physics-embedded persistent homology}
All the topological methods discussed above are  force-field-free approaches. In other words, they depend only on atomic coordinates and types without the need for molecular force field information. However, despite being insufficient, non-unique, and subject to errors, many biophysical models offer important approximations to the ground truth of biological science and reflect some of our best understandings of the biological world. Therefore, it is crucial to develop the so-called ``physics-embedded'' topology which incorporates physical models into topological invariants.

We are particularly interested in physical models that quantify the interaction strengths and directions. To characterize electrostatics interactions, we can construct a Rips filtration based on the Coulomb's potential,
\begin{equation}\label{eq:coulomb_filtration}
F_{\rm ele}(i,j) = \frac{1}{1+\exp(-cq_iq_j/d_{ij})},
\end{equation}
where the filtration value $F_{ele}(i,j)$ for the edge between atom $i$ and $j$ depends on their partial charges $q_i$ and $q_j$ and their geometric distance $d_{ij}$ \cite{ZXCang:2018a}. The part due to the Coulomb's potential in Eq.~(\ref{eq:coulomb_filtration}) can be substituted by other models, such as the van der Waals potential. We can also use cubical persistent homology \cite{Allili:2001h} to characterize the charge density as volumetric data, for example, one estimated from point charges,
\begin{equation}\label{eq:charge_density}
\mu_c(\mathbf{r}) = \sum\limits_iq_i\exp(-\|\mathbf{r}-\mathbf{r}_i\|/\eta_{i}),
\end{equation}
where $\mathbf{r}_i$ is the position of atom $i$ and $\eta_i$ is a   characteristic bond-length parameter.

 In a more general setting, there often are available properties defined on the simplices in the simplicial complex representing the protein-ligand complex. The interaction strength characterized by physical models as in Eq.~\ref{eq:coulomb_filtration} is indeed a property defined on the 1-simplices (edges). There are also various properties given on the 0-simplices (nodes/atoms) including atomic weight, atomic radii, and partial charges. Another way of incorporating these properties into the topological representation is to attach additional attributes to the persistence barcodes obtained through geometric filtration. We developed a method called \emph{enriched barcode} through cohomology theory \cite{ZXCang:2018c}. The usage of cohomology has led to efficient algorithms \cite{bauer2017ripser} as well as richer representations \cite{de2011persistent}. We are unable to elaborate on the details of cohomology here and the interested reader is referred to the aforementioned references.

Consider a persistence barcode $\{[b_i, d_i)\}_i\in I$ of dimension $k$ obtained by a geometric based filtration of the molecular system, for example, the Vietoris-Rips filtration built upon the Euclidean distance between atoms in space. Let $K(x;k)$ be the set of $k$-simplices of the simplicial complex in the corresponding filtration with the filtration parameter $x$. Our goal is to annotate each persistence pair $[b_i, d_i)$ in the barcode with the non-geometric information provided by $f:K(\infty,k)\rightarrow\mathbb{R}$. We proposed to embed such non-geometric information via cohomology \cite{ZXCang:2018c}. Specifically, for an $x\in [b_i, d_i)$, let $\omega_{i,x}$ be a real $k$-cocycle lifted from the representative cocycle from the persistent (co)homology computation \cite{de2011persistent}. A smoothed cocycle $\bar{\omega}_{i,x}=\bar{\alpha}+\omega_{i,x}$ can be obtained by solving the following problem,

\begin{equation}\label{eq:smooth_cocycle}
\bar{\alpha} = \argmin_{\alpha\in C^{k-1}(K(x), \mathbb{R})}\|\mathcal{L}(\omega_{i,x}+d\alpha)\|_2^2,
\end{equation}
where $C^{k-1}(K(x),\mathbb{R})$ is the real $(k-1)$-cochain on $K(x)$, $d$ is the coboundary operator, and $\mathcal{L}$ is an Laplacian operator. This smoothed representative $k$-cocycle $\bar{\omega}$ annotates the simplices with weights which can be used to describe the non-geometric information on this persistence pair,

\begin{equation}\label{eq:enriched_barcode}
f^*_i(x) = \sum\limits_{\sigma\in K(x;k)}f(\sigma)|\bar{\omega}_{i,x}(\sigma)|\Big/\sum\limits_{\sigma\in K(x;k)}|\bar{\omega}_{i,x}(\sigma)|.
\end{equation}
 Intuitively, this obtained function $f^*_i:[b_i,d_i)\rightarrow\mathbb{R}$ describes the average value of $f$ near the $k$-dimensional hole associated to the persistence pair $[b_i,d_i)$. We call this object enriched barcode $\{\{[b_i,d_i),f^*_i\}\}_{i\in I}$ \cite{ZXCang:2018c}. In practice, we only compute for several filtration values in the interval or even only one such as the midpoint of each persistence pair.

\subsubsection{From topological invariants to machine learning algorithms}

 While persistent homology already significantly reduces the complexity of the molecular system description, directly feeding it to machine learning algorithms can cause too many model parameters compared to the moderate size of available data in this field. Also, the outputs of persistent homology are similar to unstructured point clouds. Additional processing is needed to integrate persistent homology characterization with machine learning models.

In the application to biomolecular structure description, prior knowledge is available on the approximate distance ranges for different interactions. Therefore, we first divide an interval $[0, D]$ where $D$ is the longest range among the interactions of interest into bins. We then count the number of events in each bin, namely, 1) birth of persistence pairs, 2) death of persistence pairs, and 3) overlaps of bars with the bins. These approaches result in a 1-dimensional image-like feature tensor with three channels which can be fed into a 1-dimensional convolutional neural network or any other machine learning model that accepts structured features. Prior knowledge on the spatial range of different types of interactions can guide the decision of bin endpoints. We have also found similar performance with uniform partitioning. Another way of vectorization is to statistically describe the unstructured persistence barcodes, for example, the mean value and standard deviation of birth, death, and bar lengths.

 The Wasserstein distance between the resulting persistence barcodes also works well with distance-based methods, such as k-nearest-neighbor-based regression and classification or k-means clustering. This approach was found effective especially when the objects are moderately complex. It has been successfully applied to ligand-based tasks \cite{ZXCang:2018a}.

In general applications of integrating persistent homology with machine learning, the persistence barcodes can become sparse and available field knowledge might be insufficient to guide the vectorization. In this case, a neural network layer with each neuron learning a kernel function can automatically vectorize the barcodes. Specifically, one neuron in such layer is a function that takes the persistence barcode $\mathcal{B}=\{[b_i,d_i\}_{i\in I}$ and output a number,
\begin{equation}\label{eq:ph_neuron}
\mathcal{N}(\mathcal{B};\Theta) =  \sum\limits_{i\in I}\phi(|b_i-\mu_b|,|d_i-\mu_d|;\Theta),
\end{equation}
 where $\phi$ is a distance-based kernel function with learnable parameters $\Theta$ and the center $(\mu_b,\mu_d)$. This layer can be the first layer in a neural network for supervised learning. This layer can also be used as the first layer of an autoencoder that tries to reconstruct the persistence barcodes controlled by the Wasserstein metric. On the other hand, kernel density estimators with a fixed number of kernels can also be used as a vectorization tool. Specifically, a kernel density estimator with $n_k$ kernels each of which has $n_p$ parameters to optimize can turn a persistence barcode into a feature vector of size $n_k*n_p$. Treatment such as truncated kernels might be needed to take care of the nature of persistence barcodes that the points are only in the upper left part of the first quadrant.

\subsection{Differential geometry-based methods}\label{Method:geometry}
\subsubsection{Background}
Differential geometry has a long history of development in mathematics and has been consistently studied since the 18th century. Nowadays,  many differential geometry branches have been created from Riemannian geometry, differential topology, to Lie groups. As a result, differential geometry has been used in various indiscipline fields including physics, chemistry, economics, and computer vision. In 2005, we unfolded a curvature-based model to generate biomolecular surfaces  \cite{Wei:2005}. In the following years, we successfully formulated Laplace-Beltrami operator based minimal molecular surface (MMS) for macromolecular systems \cite{Bates:2006,Bates:2006f,Bates:2008,Bates:2009}. This approach is applied to multiscale solvation modeling in which the molecular surfaces are described via the differential geometry of surfaces. Specifically, the solute molecule is still described in microscopic detail while the solvent is treated as a macroscopic continuum to reduce a large number of degrees of freedom \cite{ZhanChen:2010a, ZhanChen:2010b, ZhanChen:2011a, ZhanChen:2012, Daily:2013, Thomas:2013}. Differential geometry-based multiscale models incorporates molecular dynamics, elasticity and fluid flow to further couple the discrete macromolecular and continuum solvent domains \cite{Wei:2009,Wei:2012, Wei:2013,  DuanChen:2012a, DuanChen:2012b}. In the past few years, we have improved the computational efficiency of the geometric modeling  by incorporating the differential geometry based multiscale paradigms in Lagrangian  \cite{XFeng:2012a, XFeng:2013b} and Eulerian representations \cite{KLXia:2014a, mu2017geometric}.

Differential geometry-based multiscale models have been used for solvation free energies prediction \cite{ZhanChen:2012,BaoWang:2015a} and ion channel transport analysis \cite{DuanChen:2012a, DuanChen:2012b, DuanChen:2013,Wei:2012,Wei:2013} to demonstrate their model efficiency in comparison with atomistic scale models.

Another type of applications of differential geometry in biomolecular systems is to utilize curvatures to characterize the macromolecular surface landscape and further infer chemical and biological properties.  For example,  the minimum and maximum curvatures are combined with the surface electrostatic potential to detect both positively charged and negatively charged protein binding sites  \cite{KLXia:2014a, mu2017geometric}.

 The other type of applications of differential geometry in molecular science is to carry out curvature-based solvation free energy prediction \cite{DDNguyen:2016c}.  In this approach, the total Gaussian, mean,  minimum, and maximum curvatures of a molecule are computed for a  molecule and correlated with its solvation free energy.

\subsubsection{Challenge}

Differential geometry based multiscale models bridge the discrete and continuum descriptions and enable physical interpretation of molecular mechanisms. Curvature-based modeling of biomolecular binding sites and solvation free energy reveals macromolecular interactive landscapes. These methods are designed as physical models to enhance our understanding of biomolecular systems.
 However, they have limited capability in predicting massive and diverse datasets due to their dependence on physical models such as the Poisson-Boltzmann equation or the Poisson-Nernst-Planck equation or their excessive reduction of geometric shape information, i.e., a molecular-level average of local curvatures.  Indeed, physical models depend on force field parameters which are subject to errors. Meanwhile, molecular-level descriptions are too coarse-gained for large datasets. In contrast, atomistic descriptions not only involve too much detail but also are not scalable for molecules with different sizes in a large dataset. As a result, machine learning algorithms cannot be effectively implemented.

To overcome these obstacles, we have designed new differential geometry-based models to extract element-level geometric information which automatically leads to scalable machine learning descriptors.  Additionally, the effort is given to encode intermolecular and intramolecular non-covalent interactions. Therefore, these novel models can be handily applied for a diverse molecular and biomolecular datasets, including protein-ligand binding analysis and prediction.

\subsubsection{Multiscale discrete-to-continuum mapping} \label{Density}

Biomolecular datasets provide atomic coordinate and type information. To facilitate differential geometry modeling, this discrete representation is transformed into a continuum one by the so-called discrete-to-continuum mapping.  In a given biomolecule or molecule with $N$ atoms, denote  $\mathbf{r}_j \in \mathbb{R}^3$ and $q_j$ the position of $j^{\rm th}$ atom and its partial charge, respectively. For any point r in three-dimensional space, a discrete-to-continuum mapping \cite{KLXia:2013d,Opron:2014,DDNguyen:2016b} defines the molecular number/charge density as the following
 \begin{align}\label{fri_surface}
	\rho(\mathbf{r}, \{ \eta_k\}, \{w_k\})=\sum_{\substack{j=1}}^{N} w_j  \Phi\left(\|\mathbf{r}-\mathbf{r}_j\|;\eta_{j}\right),
\end{align}
Especially, the density  $\rho$ indicates the molecular number density when $w_j=1$, and represents the molecular charge density when $w_j=q_j$. In addition, $\eta_j$ describes characteristic distances, $\|\cdot\|$ is the second norm, and $\Phi$  with $C^2$ property satisfies the following admissibility conditions

\begin{align}\label{eq:admiss}
\Phi \left(\|\mathbf{r}- \mathbf{r}_j\|;\eta_{j}\|\right)&=1, \quad{\rm as} \quad  \|\mathbf{r} -\mathbf{r}_j\| \rightarrow 0, \\
\Phi \left(\|\mathbf{r} - \mathbf{r}_j\|;\eta_{j}\|\right)&=0, \quad {\rm as} \quad  \|\mathbf{r} -\mathbf{r}_j\| \rightarrow \infty.
\end{align}
In principle, the density function can accept all radial basis functions (RBFs) as well as $C^2$ delta sequence of the positive type examined in this work \cite{GWei:2000}.
In practice, the generalized exponential functions
 \begin{align}\label{exponential}
\Phi\left(\|\mathbf{r}_i -\mathbf{r}_j\|;\eta_{kk'}\|\right)=e^{-\left(\|\mathbf{r}_i -\mathbf{r}_j\|/\eta_{kk'}\right)^\kappa}, \quad \kappa>0;
\end{align}
and generalized Lorentz functions
\begin{align}\label{Lorentz1}
\Phi\left(\|\mathbf{r}_i -\mathbf{r}_j\|;\eta_{kk'}\right)=\frac{1}{1+\left(\|\mathbf{r}_i -\mathbf{r}_j\|/\eta_{kk'}\right)^\nu},\quad \nu>0.
\end{align}
 seem to be the most optimal choice for the biomolecular datasets \cite{KLXia:2013d,Opron:2014}. Here power parameters $\kappa$ and $\nu$ vary for different datasets and are systemically selected.

To generate the multiscale representation for $\rho(\mathbf{r}, \{ \eta_j\}, \{w_j\})$, one can vary different values for scale parameters $\{\eta_j\}$. The published work \cite{KLXia:2015e} has shown that the molecular number density Eq. (\ref{fri_surface}) is an efficient representation for molecular surfaces. Unfortunately, such molecular-level description serves a little role in the predictive models for massive data.

\subsubsection{Element interactive densities }  \label{sec:EID}
 To handle the diversity molecular or biomolecular datasets, we have upgraded differential geometry descriptors with an emphasis on non-covalent intramolecular interactions in a molecule and intermolecular interactions in complexes, such as protein-ligand, protein-nucleic acid, and protein-protein complexes. Also, our differential geometry features can characterize the geometric information at element-specific interactions and are scalable despite a wide range of molecular sizes.

To accurately encode the physical and biological information in the differential geometry representations, we describe the molecular interactions at the element-level in a systematical manner. For instance, in the protein-ligand datasets, the intermolecular interactions are decomposed into element-level descriptions based on the commonly occurring element type in proteins and ligands. Typically, protein structures usually consist of  ${\rm H, C, N, O, S}$, and ligand structures often include ${\rm H, C, N, O, S, P, F, Cl, Br, I}$. That results in 50 element-level intermolecular descriptions. In practice, hydrogen atoms are missing in most Protein Data Bank (PDB) datasets for proteins. Therefore, we do not include it in our models for macromolecules or for both proteins and ligands. Finally, we end up with  40 or 36 element-specific groups to express the intermolecular interactions in the protein-ligand complexes. This element-specific approach can be straightforwardly carried out in other interactive systems in chemistry, biology and material science. For example, in protein-protein interactions, one can similarly arrive at a total of 16 element-level descriptions for practical use.

In a given molecule, based on the most frequently appearing element types included in the set ${\cal C}=\{{\rm H, C, N, O, S, P, F, Cl,  \cdots }\}$, we collect $N$ atoms. For each $j^{\rm th}$ atom in that collection, we label it as $\{(\mathbf{r}_j, \alpha_j,q_j)$. Here $\alpha_j$ is the element type of  $j^{\rm th}$ atom, and $\alpha_j$ = ${\cal C}_k$ indicates the $k^{\rm th}$ element type in set ${\cal C}$.

Before defining the element interactive density, we have to designate the non-covalent interactions between two element types ${\cal C}_{k}$ and ${\cal C}_{k'}$.  Such interactions can be represented by correlation kernel $\Phi$
\begin{equation}\label{CollInter}
\{\Phi(||\mathbf{r}_i-\mathbf{r}_j||; \eta_{kk'})| \alpha_i  = {\cal C}_{k}, \alpha_j  = {\cal C}_{k'};  i,j = 1,2,\ldots,N;
||\mathbf{r}_i-\mathbf{r}_j||> r_i+r_j +\sigma \},
\end{equation}
where $r_i $ and $ r_j$ are the atomic radii of $i^{\rm th}$ and $j^{\rm th}$ atoms, respectively and $\sigma$ is the mean value of the standard deviations of all $r_i $ and $ r_j$  in the dataset. The inequality constraint $||\mathbf{r}_i-\mathbf{r}_j||> r_i+r_j +\sigma $ serves the purpose of excluding the covalent forces.

Given a point ${\bf r}$ in ${\mathbb R}^3$, we define the element interactive density induced by the pairwise interaction between two chemical element types ${\cal C}_{k}$ and ${\cal C}_{k'}$
\begin{equation}\label{DG-ESRI}
\rho_{kk'}({\bf r},  \eta_{kk'}) =\sum_{j }w_j \Phi(||\mathbf{r}-\mathbf{r}_j||;\eta_{kk'}),
\quad {\bf r}  \in D_k, \alpha_j  = {\cal C}_{k'}; ||\mathbf{r}_i-\mathbf{r}_j||> r_i+r_j +\sigma, \forall \alpha_i\in    {\cal C}_{k};    k\neq k',
\end{equation}
where $D_k$ is so-called {\it atomic-radius-parametrized} van der Waals domain given by the union of all the balls with centers are the $C_k$ atomic positions with the corresponding atomic radius $r_k$. In other words, if $B({\bf r}_i,r_i)$  is denoted as a ball with a center ${\bf r}_i$ and a radius $r_i$, $D_k$ can be expressed as
\begin{equation}
D_k:= \cup_{{\bf r}_i, \alpha_i = {\cal C}_k } B({\bf r}_i, r_k).
\end{equation}

Note that element interactive density represented in (\ref{DG-ESRI}) is only good for $k\neq k'$. When density is calculated based on the interactions between the same element types, i.e. $k=k'$,  each $C_k$ atom will belong to the {\it atomic-radius-parametrized } van der Waals domain and element interactive density representation. To this end, we define such density formulation as the following
\begin{equation}\label{DG-ESRI2}
\rho_{kk}({\bf r}, \eta_{kk}) =  \sum_{j } w_j \Phi(||\mathbf{r}-\mathbf{r}_j||; \eta_{kk}),
\quad {\bf r}  \in D^i_k,   \alpha_i  = {\cal C}_{k};  \alpha_j  = {\cal C}_{k}; ||\mathbf{r}_i-\mathbf{r}_j||> 2r_j +\sigma,
\end{equation}
in which, domain $D^i_k$ is just a single ball $B({\bf r}_i,r_i)$, and the density function $\rho_{kk}$ is evaluated at all $D^i_k$.

 The element interactive density $\rho_{kk}$ is the linear combination of correlation kernel $\Phi$ of pairs of element types. Consequently, the smoothness of $\rho_{kk}$ is the same as that of  $\Phi$. Moreover, by changing a level constant $c$, one can attain a family of element interactive manifolds as

 \begin{equation}\label{manifold}
 \rho_{kk'}({\bf r},  \eta_{kk'})=c\rho_{\max}, \quad 0\leq c \leq 1 \quad {\rm and } \quad \rho_{\max}=\max\{\rho_{kk'}({\bf r},  \eta_{kk'})\}.
\end{equation}
Figure \ref{fig:DG-flowchart} illustrates a few element interactive manifolds.

\subsubsection{Element interactive curvatures}

\paragraph{Differential geometry of differentiable manifolds}
We here describe the geometric information calculation on a differential manifold. Consider $U$ being an open subset of ${\mathbb R}^n$ with its closure is compact \cite{Wolfgang:2002,Bates:2008,Wei:2009}, we are interested in a $C^2$ immersion ${\bf f}: U\rightarrow {\mathbb R}^{n+1}$. Given a vector ${\bf u} = (u_1,u_2,\cdots,u_n) \in U$, we express the Jacobian matrix w.r.t ${\bf u}$ as
\begin{equation}
D{\bf f}=(X_1,X_2,\cdots,X_n), \quad X_i=\frac{\partial {\bf f}} {\partial u_i},i=1,2\cdots n.
\end{equation}
The first fundamental form is written in the metric tensor with its coefficients $g_{ij}=\left\langle
X_i,X_j\right\rangle$, where $\left\langle,\right\rangle$ is the
Euclidean inner product in ${\mathbb R}^n$, $i,j =1,2,\cdots,n$.

We define the unit normal vector via the Gauss map
\begin{alignat}{2}
  {\bf N}: U&\longrightarrow& R^{n+1} &\\
  (u_1,u_2,\cdots,u_n)&\longmapsto& X_1 \times & X_2 \cdots \times X_n/ \|X_1
\times X_2 \cdots \times X_n\|,
\end{alignat}
where $``\times'' $ denotes the cross product. If we denote $\bot_{\bf
u}{\bf f}$  the normal space of ${\bf f}$ at point ${\bf X}={\bf
f(u)}$, then ${\bf N}({\bf u}) \in \bot_{\bf u} {\bf f}$. In addition, one can form a second fundamental form via the means of the normal vector ${\bf N}$ and tangent vector $X_i$:
\begin{equation}
II(X_i,X_j)=(h_{ij})_{i,j=1,2,\cdots
n}=\left(\left\langle-\frac{\partial{\bf N}}{\partial u_i},
X_j\right\rangle\right)_{ij}.
\end{equation}
Then, the Gaussian curvature $K$ and the mean curvature $H$ are determined as the following
\begin{align} \label{fundamental_curvatures}
K=\frac{{\rm Det}(h_{ij})}{{\rm Det}(g_{ij})}, \quad H=\frac{1}{n}h_{ij}g^{ji}.
\end{align}
The Einstein summation convention is used in the curvature expressions and $(g^{ij})=(g_{ij})^{-1}$.

\paragraph{Element interactive curvatures}
With an element interactive manifolds defined via element interactive density $\rho({\bf r})$ describing in (\ref{manifold}) and the expressions in (\ref{fundamental_curvatures}),  one can further formulate the representations for the Gaussian curvature ($K$)  and the mean curvature ($H$) as the following \cite{Soldea:2006,KLXia:2014a}
    \begin{align}
    K=&\frac{1}{g^2}\left[
    2\rho_x \rho_y \rho_{xz}\rho_{yz} + 2\rho_x\rho_z\rho_{xy}\rho_{yz}+2\rho_y\rho_z\rho_{xy}\rho_{xz} \right. \nonumber\\
    &\left.  - 2 \rho_x \rho_z \rho_{xz} \rho_{yy} - 2 \rho_y \rho_z \rho_{xx} \rho_{yz} - 2 \rho_x \rho_y \rho_{xy} \rho_{zz} \right.\nonumber\\
    &\left. +\rho_z^2 \rho_{xx}  \rho_{yy} + \rho_x^2 \rho_{yy} \rho_{zz} + \rho_y^2 \rho_{xx} \rho_{zz}\right.\nonumber\\
    &\left. -\rho_x^2 \rho_{yz}^2 - \rho_y^2 \rho_{xz}^2 - \rho_z^2 \rho_{xy}^2 \right],
    \label{gaussian_curv}
    \end{align}
    and
    \begin{align}
    H=\frac{1}{2g^{\frac{3}{2}}}\left[
    2 \rho_x \rho_y \rho_{xy} + 2 \rho_x \rho_z \rho_{xz} + 2 \rho_y \rho_z \rho_{yz} - (\rho_y^2 + \rho_z^2)\rho_{xx} - (\rho_x^2 + \rho_z^2)\rho_{yy} - (\rho_x^2+\rho_y^2)\rho_{zz}\right],
    \label{mean_curv}
    \end{align}
    where $g=\rho_x^2 + \rho_y^2 + \rho_z^2$.

    In addition, the minimum curvature ($\kappa_{\rm min}$)  and maximum curvatures ($\kappa_{\rm max}$) can be evaluated based on the Gaussian and mean curvature values
    \begin{align}
    \kappa_{\rm min}=H-\sqrt{H^2-K},\quad \kappa_{\rm max}=H+\sqrt{H^2-K}.\label{minmax_curv}
    \end{align}

It is noted that in the curvature representations in (\ref{gaussian_curv}), (\ref{mean_curv}), and (\ref{minmax_curv}), the derivatives of the density function can be analytically calculated. For the convenience, we denote the curvatures associated with the density function $\rho_{kk'}({\bf r}, \eta_{kk'})$ as $K_{kk'}({\bf r}, \eta_{kk'})$, $H_{kk'}({\bf r}, \eta_{kk'})$, $\kappa_{kk',\rm min}({\bf r}$, $\eta_{kk'})$, $\kappa_{kk',\rm max}({\bf r}$, $\eta_{kk'})$. In practical use, the element interactive curves are only evaluated at the atomic positions in a given molecule or biomolecule structure. To achieve element-level geometry information, we propose the element interactive mean curvature as the following

\begin{equation}\label{InterCurv}
   H^{\rm EI}_{kk'}(\eta_{kk'}) = \sum_i  H_{kk'}({\bf r}_i, \eta_{kk'}), \quad {\bf r}_i\in D_k; k\neq k'
\end{equation}
and
\begin{equation}\label{InterCurv2}
   H^{\rm EI}_{kk}(\eta_{kk}) =   \sum_i  H_{kk}({\bf r}_i, \eta_{kk}), \quad {\bf r}_i\in D^i_k, D^i_k \subset D_k.
\end{equation}

The other element-level interactive curvatures for Gaussian curvature ($ K^{\rm EI}_{kk'}(\eta_{kk'})$), minimum curvature ($\kappa^{\rm EI}_{kk',\rm min}(\eta_{kk'})$), and maximum curvature ($ \kappa^{\rm EI}_{kk',\rm max}( \eta_{kk'})$) are defined in a similar manner.

\subsubsection{Differential geometry  based geometric learning (DG-GL)} \label{sec:DG-GL_learning}
\paragraph{Geometric learning}
\begin{figure}[!ht]
    \centering
     \includegraphics[width= 1 \textwidth]{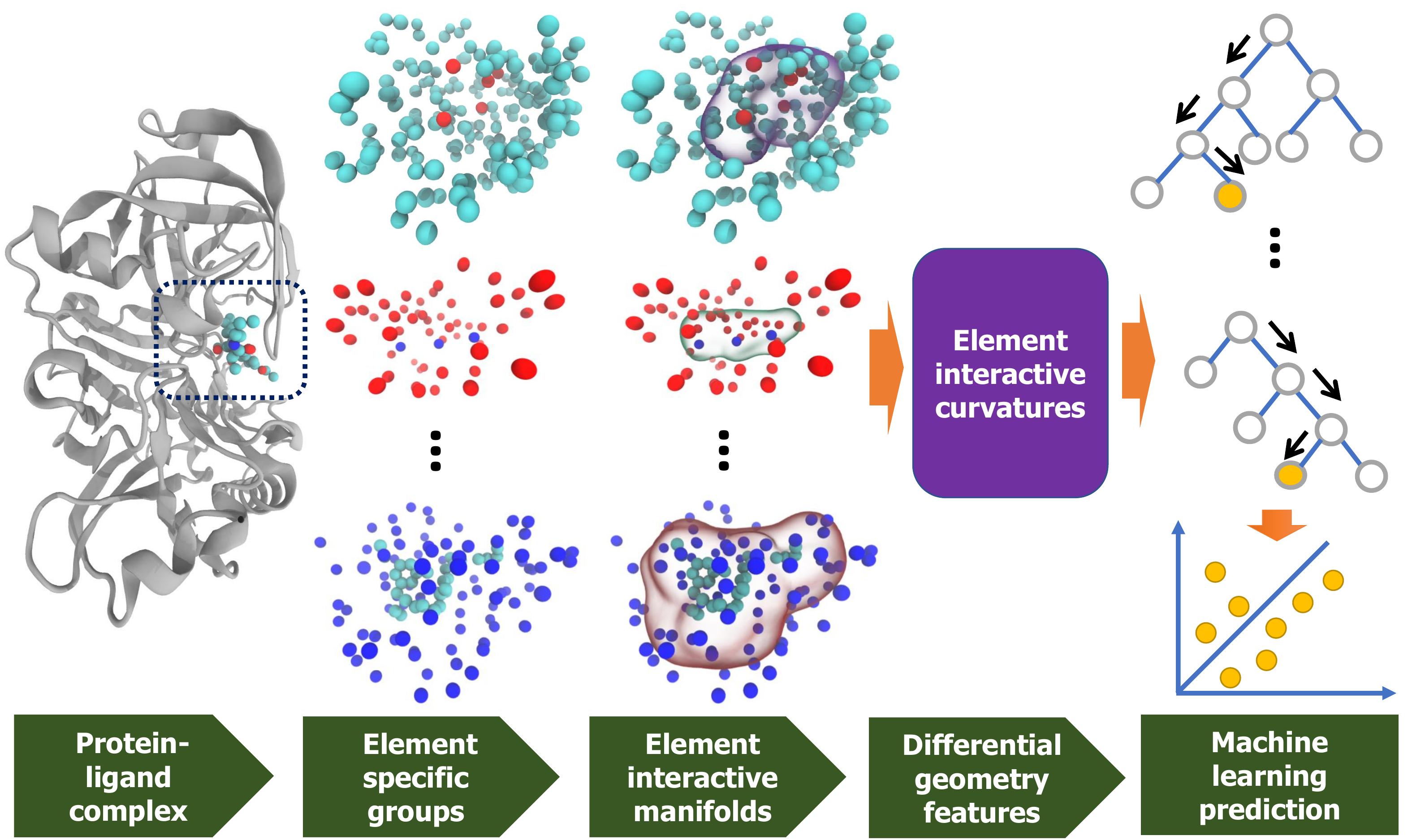}
    \caption{ Illustration of  the DG-GL strategy for complex with PDBID: 5QCT (first column).  The second column presents the different element specific groups including OC, CN, and CH, respectively from top to bottom. The third column depicts the element interactive manifolds for the corresponding element specific groups. A predictive model in the last column integrates the differential geometry features (fourth column) with the machine learning algorithm.}
    \label{fig:DG-flowchart}
\end{figure}
In our differential geometry based geometric learning (DG-GL) model, we incorporate the geometric representations such as element-level interactive curvatures with advanced machine learning algorithms to form powerful predictive models. Given a training set $\{{\cal X}_i\}_{i\in I}$, in which ${\cal X}_i$ is the input data for the $i^{\rm th}$ molecule and $I$ is the set of the molecular indices in the training data. We denote $\mathbf{F}({\cal X}_i; \bm{\zeta})$ is a differential geometric functions encoding the the input structures ${\cal X}_i$ via the given hyperparameter set ${\bm \zeta}$ into aforementioned DG descriptions. Our DG-GL model learns the training set $\{{\cal X}_i\}_{i\in I}$ by minimizing the following loss functions
\begin{equation}
\min\limits_{\bm{\zeta}, \bm{\theta}} \sum\limits_{i\in I}L(\bf{y}_i, \mathbf{F}({\cal X}_i; {\bm \zeta}); {\bm \theta}),
\end{equation}
in which $L$ is the loss function, ${\bf y}_i$ is the target label of molecule ${\cal X}_i$, and ${\bm \theta}$ is the set of parameters of a selected machine learning algorithm. It is worth noting that the DG representation encoded in ${\mathbf F}$ does not depend on the type of learning task. Therefore, our DG-GL models can adapt any regressors or classifiers models such as linear regression, support vector machine,  random forest, gradient boosting trees, artificial neural networks, and convolutional neural networks. Besides the machine learning hyperparameters, the kernel parameters in the encoding DG function ${\mathbf F}$ need to be optimized for a specific learning algorithm and a particular training set $\{{\cal X}_i\}$.

 In the validation, we only utilize the gradient boosting trees (GBTs) even though the other advanced machine learning models including convolutional neural networks can be incorporated with minimal effort. The general framework of DG-GL model is depicted in (\ref{fig:DG-flowchart}). The GBTs in the DG-GL score are employed via the gradient boosting regression module in scikit-learn v0.19.1 package with the following hyperparameters: n\_estimators=10000, max\_depth=7, min\_samples\_split=3, learning\_rate=0.01, loss=ls, subsample=0.3, max\_features=sqrt for all experiments.

\paragraph{Model parametrization}
In our differential geometry-based approach, we calculate the element interactive curvatures (EICs)  of type $C$ based on kernel $\alpha$ with parameters ($\beta, \tau$). We denote such model ${\rm EIC}^{C}_{\alpha, \beta, \tau}$ . Here, $C \in \{K, H, k_{\min}, k_{\max}\}$ and  $\alpha={\rm E}$ and $\alpha={\rm L}$ indicate generalized exponential and generalized Lorentz kernels, respectively. In addition, $\beta$ refers to the kernel order and is denoted as $\kappa$ if $\alpha={\rm E}$ or $\nu$ if $\alpha={\rm L}$. Another kernel parameter is $\tau$  defined by the following relationship
\begin{equation}
\eta_{kk'}=\tau (\bar{r}_k + \bar{r}_{k'})
\end{equation}
where $\bar{r}_k$ and $\bar{r}_{k'}$ stand for the van der Waals radii of element type  $k$ and element type $k'$, respectively. These kernel parameters are selected via a 5-fold cross-validation on a specific training set with the range of $\tau$ and $\beta$ varying from 0.5 to 6 with an increment of 0.5. Moreover, we are interested in high values of power order, $\beta$ $\in\{10, 15, 20\}$ ,  which accounts for the ideal low-pass filter (ILF)  \cite{KLXia:2015f}. These parameter ranges are also listed in Table \ref{tab:DG-GL_parameter_domains}.

\begin{table}[!ht]
    \centering
    \caption{The ranges of DG-GL hyperparameters for 5-fold cross-validations}
        \label{tab:DG-GL_parameter_domains}
    \begin{tabular}{|c|c|}
        \hline
        Parameter & Domain\\ \hline
        $\tau$  & $\{0.5, 1.0, \dots, 6\}$ \\\hline
        $\beta$ & $\{0.5, 1.0, \dots, 6\} \cup \{10, 15, 20\}$\\\hline
        $C$ & $\{K, H, k_{\min}, k_{\max}\}$\\
        \hline
    \end{tabular}
\end{table}

To enable the multiscale descriptions in  differential geometry representation, we employ multiple kernels to evaluate the EICs. For instance, if two kernels with the following parameters $(\alpha_1, \beta_1,\tau_1)$ and $(\alpha_2,\beta_2,\tau_2)$ are utilized, our EIC model can be written as ${\rm EIC}^{C_1C_2}_{\alpha_1, \beta_1,\tau_1;\alpha_2,\beta_2,\tau_2}$.

 In a protein-ligand complex, we are interested in 4 commonly occurred protein atom types \{${\rm C, N, O, S}$\}, and 10 commonly occurred ligand atom types \{${\rm H, C, N, O, F, P, S, Cl, Br, I}$\}. That results in a total of 40 different combinations. With a set of calculated atomic pairwise curvatures, we construct 10 statistical features, namely sum, the sum of absolute values, minimum, the minimum of absolute values,  maximum, the maximum of absolute values, mean, the mean of absolute values, standard deviation, and the standard deviation of absolute values. In total, we attain 400 features for the current differential geometry-based models.

\subsection{Graph theory-based methods}\label{Method:graph}

\subsubsection{Background}
Graph theory is one of the most popular subjects in discrete mathematics. In graph theory, the information inputs are represented in the graph structures formed by vertices that are connected by edges and/or high-dimensional simplexes. Different ways to interpret the graph result in different graph theories such as geometric graph theory, algebraic graph theory, and topological graph theory. In geometric graph study, the graph information is extracted based on the geometric objects drawn in the Euclidean plane \cite{pach2013beginnings}. If there are algebraic methods involving in graph structure processing, that approach belongs to algebraic graph theory. There are two common approaches to this branch. The first one is to use linear algebra to study the spectrum of various types of matrices representing graph including adjacency matrix and Laplacian matrix \cite{godsil2013algebraic}. Another approach relies on the group theory, especially automorphism groups \cite{babai1996automorphism} and geometric group theory \cite{de2000topics}, for the study of graphs. Unlike the aforementioned graph theories, the algebraic graph theory considers graphs as topological spaces by associating different types of simplicial complexes such as abstract simplicial complex \cite{korte2012greedoids} and Whitney complex \cite{larrion2002whitney}.

Due to the natural representations for structured information, graph theory enacts enormous applications in various fields including computer science, linguistics, physics, chemistry, biology, and social sciences.
 Especially in the chemical and biological study, graph theory is commonly used since molecular structures always feature graph information in which vertices illustrate atoms and graph edges represent bonds. Indeed, graph-based approaches have been utilized to describe chemical datasets \cite{balaban1976chemical,trinajstic1983chemical, schultz1989topological, foulds2012graph, hansen1988chemical, ozkanlar2014chemnetworks} as well as biomolecular datasets \cite{Bahar:1997,di2015protein, canutescu2003graph, ryslik2014graph,Jacobs:2001, vishveshwara2002protein,wu2017moleculenet}. In addition, one can make use of graph representations to uncover the connectivity of different components of a molecule  such as  centrality \cite{newman2010networks,bavelas1950communication,dekker2005conceptual}, contact map \cite{Bahar:1997, LWYang:2008},  and topological index \cite{hosoya1971topological,hansen1988chemical}. Moreover, graph extracting descriptors can be employed in chemical analysis \cite{trinajstic1983chemical,schultz1989topological,  janezic2015graph} and biomolecular modeling \cite{Angeleska:2009}. Particularly, some research groups have invested their efforts to carry out the graph-based representation to model protein flexibility and long-time dynamics such as normal-mode analysis (NMA) \cite{Go:1983,Tasumi:1982,Brooks:1983,Levitt:1985}  and elastic network model (ENM) \cite{Flory:1976, Bahar:1997,Bahar:1998,Atilgan:2001,Hinsen:1998,Tama:2001}.

\subsubsection{Challenge}

Due to the rich in geometric interpretations, graph theory-based approaches have shown their efficiency in the qualitative and descriptive models. However, oversimplified representations and the lack of physical and biological detailed information may render graph theory-based approaches less attractive in the quantitative analysis. For instance, in Gaussian network model (GNM) \cite{Bahar:1997,Bahar:1998,QCui:2010}, the use of the spectrum of the Laplacian matrix is quite efficient to decompose the flexible and rigid regions and domains of proteins but its fluctuation predictions on protein C$_\alpha$ atoms were not reliable with the Pearson correlation coefficient as low as 0.6 for three datasets \cite{JKPark:2013}. To predict the mutations in proteins, the graph-based  mCSM method was not competent as physical and knowledge-based or topological fingerprint-based models \cite{LQuan:2016,ZXCang:2017a}.

The poor performances of the aforementioned graph theory-based models on   quantitative tasks are due to the lack of three main components in our point of view. Firstly, these graph theory-based structures do not provide the information at the chemical element   level.  Consequently, these models treat different element types equally which results in inadequate coded information from the original structures. Secondly, non-covalent interactions between the two atoms are overlooked in many graph edges which cause the unphysical representations for most molecular and biomolecular data. Finally, the edges in the many graph-based models express the connectivity between a pair of atoms based on the number of covalent bonds between these two atoms, which inaccurately describe  many interactions that depend on the Euclidean distance.

 To address the aforementioned issues in graph based-modeling, we have developed the weighted graphs, termed as the flexibility-rigidity index (FRI), to predict the B-factor of protein atoms.  In our FRI model, the graph edges were formulated by the radial basis functions (RBFs) \cite{KLXia:2017, Opron:2014, Opron:2015a, DDNguyen:2016b} which properly describe the interaction strengths between two atoms in the equilibrium structures. The original FRI was upgraded to multiscale FRI \cite{Opron:2015a, KLXia:2015f} for capturing the multiscale interactions in biological structures. Specifically, the graph in the multiscale FRI model is allowed to have multiple edges formed by RBFs with careful selections of scaled and power parameters. Although our FRI models have outperformed the GNM in B-factor predictions, they provide only coarse-grained molecular-level descriptions.  To overcome this limitation, we have proposed  graph coloring based methods with vertices colored differently based on the corresponding element types. Consequently, we ended up with various element-specific subgraphs taking care of different types of physical interactions, such as hydrophilic, hydrophobic, hydrogen bonds \cite{DDNguyen:2017d, DBramer:2018a}.  As a result, the predicted accuracy for protein B-factors by our multiscale weighted colored graphs is over 40\% higher than GNM models \cite{DBramer:2018a}. The success of multiscale weighted colored graph models on B-factor prediction encouraged us to design graph-based scoring functions to predict protein-ligand binding affinities. The protein-ligand binding mechanism is more complex than the protein B-factor. Therefore, it requires sophisticated graph-based models to accurately encode the physical and biological properties to unveil its molecular mechanism. The development of such graphs is described in the following sections.

\subsubsection{Multiscale weighted colored geometric subgraphs}

 In this section, we discuss general graph representations for a molecule or biomolecule. Graph-based descriptors are systematical, scalable, and straightforward applied not only to the predictions of protein-ligand binding affinity but also for various bioactivities such as toxicity, solvation, solubility, partition coefficient, mutation-induced protein folding stability change, and protein-nucleic acid interactions. In a given molecule or biomolecule in a dataset, we denote a graph ${\cal G}$ to represent a subset of its $N$ atoms. The set of its vertices  ${\cal V}$ consists of  coordinates and chemical element types of atoms, defined as
\begin{equation}
{\mathcal V} = \{(\mathbf{r}_j, \alpha_j)|\mathbf{r}_j\in {\rm I\!R}^3; \alpha_j \in {\mathcal C};  j=1,2,\ldots,N \},
\label{eq:vertices}
\end{equation}
where $\mathbf{r}_j$ is the 3D position of $j^{\rm th}$ atom, and $\alpha_j$ is its element type which belongs to a predefined set of commonly occurred chemical element types as introduced in Section \ref{sec:EID}.
To accomplish a meaningful encoded physical and biological information in the graph, graph edges have to express the non-covalent interactions. Moreover, to accommodate for the interactions between $k$ element atoms and $k'$ element type atoms, we consider a set of graph edges ${\cal E}_{kk'}$ represented by RBFs as the following
\begin{multline}\label{AGL_CollInter}
{\mathcal E}_{kk'}= \{\Phi(||\mathbf{r}_i-\mathbf{r}_j||; \eta_{kk'})| \alpha_i  = {\mathcal C}_{k}, \alpha_j  = {\mathcal C}_{k'};  i,j = 1,2,\ldots,N;
||\mathbf{r}_i-\mathbf{r}_j||> r_i+r_j +\sigma \},
\end{multline}
where $||\mathbf{r}_i-\mathbf{r}_j||$ accounts for the Euclidean distance between the $i^{\rm th}$ and $j^{\rm th}$ atoms, $r_i $ and $ r_j$ are the atomic radii of $i^{\rm th}$ and $j^{\rm th}$ atoms, respectively. Moreover, $\sigma$ is the mean value of the standard deviations of all atomic radii belonging to element types ${\cal C}_k$ and ${\cal C}_{k'}$  in the dataset. The exclusion of the covalent interactions are portrayed in this inequality $||\mathbf{r}_i-\mathbf{r}_j||> r_i+r_j +\sigma $. $\Phi$ is a predefined RBF representing a graph weight and has the following properties \cite{KLXia:2013d,Opron:2014}
\begin{align}
\Phi(||\mathbf{r}_i-\mathbf{r}_j||;\eta_{kk'})&=1, \mbox{ as } ||\mathbf{r}_i-\mathbf{r}_j||\rightarrow 0 \quad {\rm and }\\
\Phi(||\mathbf{r}_i-\mathbf{r}_j||;\eta_{kk'})&=0 \mbox{ as } ||\mathbf{r}_i-\mathbf{r}_j||\rightarrow\infty,
\quad \alpha_i  = {\mathcal C}_{k}, \alpha_j  = {\mathcal C}_{k'},
 \end{align}
 where $\eta_{kk'}$ is a characteristic distance between the atoms. We now achieve the weight colored subgraphs (WCS) ${\cal G(V, E}_{kk'})$ or denote ${\cal G}_{kk'}$ for short.

 In principle, our WCS ${\cal G(V, E}_{kk'})$ can adopt any RBFs. In practice, the generalized exponential functions (\ref{exponential})
and generalized Lorentz functions (\ref{Lorentz1})
 seem to be the most optimal choice for the biomolecular datasets \cite{KLXia:2013d,Opron:2014}. Here power parameters $\kappa$ and $\nu$ vary for different datasets and are systemically selected.
\begin{figure}[!htpb]
\centerline{\includegraphics[keepaspectratio,width=6.5in]{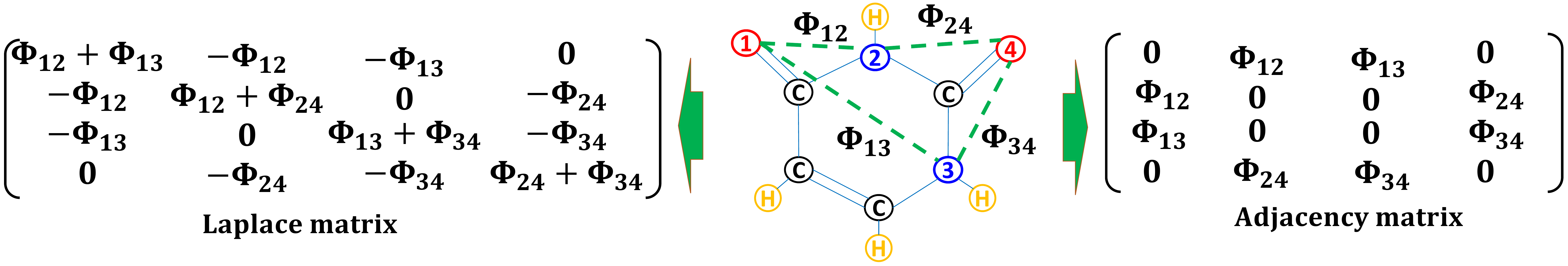}}
\caption{Illustration of   weighted colored subgraph ${\cal G}_{\rm NO}$ (Left), {  its  Laplacian matrix (Middle), and adjacency matrix (Right)} for uracil molecule (C$_4$H$_4$N$_2$O$_2$).  Graph vertices, namely oxygen (i.e., atoms 1 and 4)  and nitrogen (i.e., atoms 2 and 3), are labeled in red and blue colors, respectively. Here,  graph edges (i.e., $\Phi_{ij}$) are labeled by green-dashed lines  which are {\it not} covalent bonds. {   Here, $\Phi_{ij}$ are  distance-weighted edges.}  Note that there are 9 other nontrivial subgraphs for this molecule (i.e., ${\cal G}_{\rm CC}, {\cal G}_{\rm CN}, {\cal G}_{\rm CO}, {\cal G}_{\rm CH}, {\cal G}_{\rm NN}, {\cal G}_{\rm NH}, {\cal G}_{\rm OO}, {\cal G}_{\rm OH}, {\cal G}_{\rm HH}$).
}
\label{fig:Graph}
\end{figure}
To illustrate WCS of a given molecule, we use the uracil compound (C$_4$H$_4$N$_2$O$_2$) as an example. Figure \ref{fig:Graph} depicts WCS for nitrogen and oxygen atoms (${\cal G}_{\rm NO}$). To elicit the geometrical invariants of WCS formed by element types  ${\cal C}_{k}$ and ${\cal C}_{k'}$, we propose a collective descriptor at the element level as follows
\begin{equation}\label{ESRI}
 {\rm RI}^G(\eta_{kk'}) =\sum_i \mu^G_i(\eta_{kk'})=\sum_i\sum_{j } \Phi(||\mathbf{r}_i-\mathbf{r}_j||;\eta_{kk'}),
\quad \alpha_i  = {\cal C}_{k}, \alpha_j  = {\cal C}_{k'};  ||\mathbf{r}_i-\mathbf{r}_j||> r_i+r_j +\sigma,
\end{equation}
where  $\mu^G_i(\eta_{kk'})$ which is a geometric subgraph centrality for the $i^{\rm th}$ atom has been developed in our previous work for protein B-factors predictions \cite{DBramer:2018a}. The summation over  $\mu^G_i(\eta_{kk'})$  in Eq. (\ref{ESRI}) gives rise to WCS rigidity between  element types  ${\cal C}_{k}$ and ${\cal C}_{k'}$. In fact, $\mu^G_i(\eta_{kk'})$ is the generalized form of our successful rigidity index model for protein-ligand binding affinity prediction in the previous work \cite{DDNguyen:2017d}. it is noticed that the WCS for the protein-ligand system is bipartite since each of its edges presents the interaction between one atom in the protein and another protein in the ligand. With that design, a variety of physical and biological properties such as electrostatics, van der Waals interactions,  hydrogen bonds, polarization,  hydrophilicity,  hydrophobicity can be successfully encoded in our WCS representations.

To exhibit the intermolecular and intramolecular properties, one can vary the characteristic distance  $\eta_{kk'}$ to set up multiscale weighted colored subgraphs (MWCS). To methodically attain multiscale graph-based molecular and biomolecular descriptors in a collective and scalable manner, one can aptly select groups of pairwise element interactions $k$ and $k'$, the choice of subgraph weights $\Phi$ and their parameters.

\subsubsection{Multiscale weighted colored algebraic subgraphs}
In this section, we present another approach to extract the meaningful descriptor for biomolecules from their WCS. This scheme depends on the algebraic graph or spectral graph formulations. Since geometric and algebraic approaches handle the graph information differently. Therefore, these two kinds of subgraphs will be expected to encode the physical and biological information in varied aspects. In the algebraic graph theory, matrices are utilized to represent a given subgraph.  Two of the most common ones are the Laplacian matrix and the adjacency matrix.

\paragraph{Multiscale weighted colored Laplacian matrix }
Considering a weighted colored subgraph ${\cal G(V,E}_{kk'})$ defined at Eqs. (\ref{eq:vertices}) and (\ref{AGL_CollInter}), we construct a following weighted colored Laplacian matrix $L(\eta_{kk'})=(L_{ij}(\eta_{kk'}))$ describing the interaction between element types ${\cal C}_k$ and ${\cal C}_{k'}$
\begin{equation} \label{Laplacianmatrix}
L_{ij}(\eta_{kk'}) = \left\{ \begin{array}{ll}
     - \Phi(||\mathbf{r}_i - \mathbf{r}_j||; \eta_{kk'}) &
     \begin{aligned}
        &\text{if}~  i\neq j, \alpha_i  = {\mathcal C}_{k}, \alpha_j  = {\mathcal C}_{k'}\\
        &\text{and}~  ||\mathbf{r}_i-\mathbf{r}_j||> r_i+r_j +\sigma ;
     \end{aligned}
        \\
     -\sum_j L_{ij}  &  {\rm if}~~ i=j.
        \end{array} \right.
\end{equation}
For the illustration, we explicitly formulate the Laplacian matrix of the WCS ${\cal G}_{\rm NO}$ for the uracil molecule (C$_4$H$_4$N$_2$O$_2$) in Figure \ref{fig:Graph}. It is obvious to learn that all eigenvalues of our element-level WCS Laplacian matrix are nonnegative due to its symmetric, diagonally dominant, and positive-semidefinite properties. Moreover, every row sum and column sum of $L(\eta_{kk'})$ is zero. In consequence, its first eigenvalue is $0$. The second smallest eigenvalue of $L(\eta_{kk'})$ is so-called algebraic connectivity (also known as Fiedler value) which approximates the sparest cut of a graph. With a given WCS ${\cal G(V,E}_{kk'})$ one can easily see its geometrical invariant proposed at Eq. (\ref{ESRI}) is fully recovered in the trace of its Laplacian matrix $L(\eta_{kk'})$
\begin{equation}
{\rm RI}^G(\eta_{kk'})={\rm Tr} L(\eta_{kk'}),
\end{equation}
where ${\rm Tr}$ is the trace.

In the algebraic graph, we are interested in using the eigenvalue and eigenvector information to extract the graph invariants. To this end, we denote $\lambda^L_j, j=1,2,\cdots$ and ${\bf u}^L_j, j=1,2,\cdots$ the eigenvalues and eigenvectors of  $L(\eta_{kk'})$. The element-level molecular descriptors of the Laplacian matrix $L(\eta_{kk'})$ is proposed as the following
\begin{equation}
{\rm RI}^L(\eta_{kk'}) =\sum_i \mu^L_i(\eta_{kk'}),
\end{equation}
where $\mu^L_i(\eta_{kk'})$ is so-called an atomic descriptor for the $i^{\rm th}$ atom (${\bf r}_i, \alpha_i={\cal C}_k$)
\begin{equation}\label{eq:atomic}
\mu^L_i(\eta_{kk'})=\sum_l (\lambda^L_l)^{-1}\left[{\bf u}^L_l({\bf u}^L_l)^T \right]_{ii},
\end{equation}
 where $T$ is the transpose. It is noted that $\mu^L_i(\eta_{kk'})$ is the atomic descriptor of the generalized GNM \cite{Bahar:1997, KLXia:2015f}. Therefore, it can be directly utilized to capture atomic properties such as protein B-factors. Moreover, the element-level invariant of the Laplacian matrix can be enriched via the statistical information of $\mu^L_i(\eta_{kk'})$ values, namely sum, mean, maximum, minimum and standard deviation.

Another way to extract the invariant descriptor from the WCS Laplacian matrix is the direct use of nontrivial eigenvalues $\{\lambda^L_j\}_{j=2,3,\cdots}$. Also, the statistical analysis of those eigenvalues can be incorporated to form a feature vector to characterize element-level information of the molecule and biomolecule.

\paragraph{Multiscale weighted colored  adjacency matrix }
By setting all diagonal entities of the Laplacian matrix to be 0, we end up with an adjacency matrix with simpler representation but still preserve the essential properties of the original molecular structures. With a given WCS ${\cal G}_{kk'}$, the adjacency matrix $A(\eta_{kk'})=(A_{ij}(\eta_{kk'}))$ is given as
\begin{equation} \label{adjacencymatrix}
A_{ij}(\eta_{kk'}) = \left\{ \begin{array}{ll}
     \Phi(||\mathbf{r}_i - \mathbf{r}_j||; \eta_{kk'}) &
    \begin{aligned}
        &\text{if}~  i\neq j, \alpha_i  = {\mathcal C}_{k}, \alpha_j  = {\mathcal C}_{k'}\\
        &\text{and}~  ||\mathbf{r}_i-\mathbf{r}_j||> r_i+r_j +\sigma;
    \end{aligned}
        \\
   0 &  {\rm if}~~ i=j .
        \end{array} \right.
\end{equation}
Since the adjacency matrix defined in (\ref{adjacencymatrix}) is undirected, $A(\eta_{kk'})$ is symmetric. Thus, all the eigenvalues of it are real. Moreover, due to being a bipartite graph, for each eigenvalue $\lambda$, its opposite $-\lambda$ is also an eigenvalue of $A(\eta_{kk'})$. In consequence, only positive eigenvalues are used in the molecular descriptor. For the sake of illustration, Figure \ref{fig:Graph} illustrates the adjacency matrices for the weighted colored subgraph $G_{\rm NO}$ in the uracil molecule (C$_4$H$_4$N$_2$O$_2$). It can be seen from the Perron-Frobenius theorem that the spectral radius of $A(\eta_{kk'})$, denoted as $\rho(A)$, is bounded by the range of the diagonal elements of the corresponding Laplacian matrix
\begin{equation}
\min_i\sum_j A_{ij} \leq \rho(A)\leq \max_{i} \sum_j A_{ij}.
\end{equation}
It is easy to see that all elements in the Laplacian matrix belong to [0,1] and depends on the scale parameter $\eta_{kk'}$. At a characteristic scale range for capturing hydrogen bonds or van der  Waals interactions, the Laplacian matrix has many zeros.  However, the scale parameter $\eta_{kk'}$ can be very huge in electrostatic and hydrophobic interactions \cite{ZXCang:2017b}, which results in many elements in the Laplacian matrix nearly 1. In that particular situation, the spectral radius of  the adjacency matrix $A(\eta_{kk'})$ is bounded by $n-1$, where $n$ is the number of atoms in WCS ${\cal G}_{kk'}$.

Similarly to the approach of forming feature representation for the Laplacian matrix, all positive eigenvalues $\{\lambda^A_j\}$, and their statistical information such as sum, mean, maximum, minimum, and standard deviation are included in element-level molecular descriptors. If we define $\{{\bf u}^A_j\}$ as the eigenvectors corresponding to eigenvalues $\{\lambda^A_j\}$, then the atomic descriptors can be attained as
\begin{equation}\label{eq:atomicA}
\mu^A_i(\eta_{kk'})= \sum_ j  \left[Q \Lambda Q^{-1}\right]_{ij},
\end{equation}
where $Q=[{\bf u}^A_1  {\bf u}^A_2\cdots {\bf u}^A_n]$ is composed by $n$ linearly independent eigenvectors of $A(\eta_{kk'})$; thus $Q$ is invertible. Moreover, $\Lambda$ is a diagonal matrix with each diagonal element $\Lambda_{ii}$ being the eigenvalue $\{\lambda^A_i\}$. Unfortunately, formulation given in Eq. (\ref{eq:atomicA}) is very computationally expensive due the involvement of the inverse-matrix calculation.

In general, the methods regarding the eigenvalues and eigenvectors analysis often pose a great challenge for sustaining an efficient computation strategy. Fortunately, the construction of WCS enables us to design a less-expensive computational model due to two facts. Firstly, the protein-ligand binding site only involves a small region of the whole complex structure. Second, WCS only admits the specific element types in the matrix construction, which further reduces the size of matrices for eigenvalue and eigenvector calculations. As a result, these facts offer an efficient spectral graph-based model for protein-ligand affinity analysis.

\subsubsection{Graph-based learning models}

\paragraph{Graph learning}
\begin{figure}[!ht]
    \centering
     \includegraphics[width=1.0\textwidth]{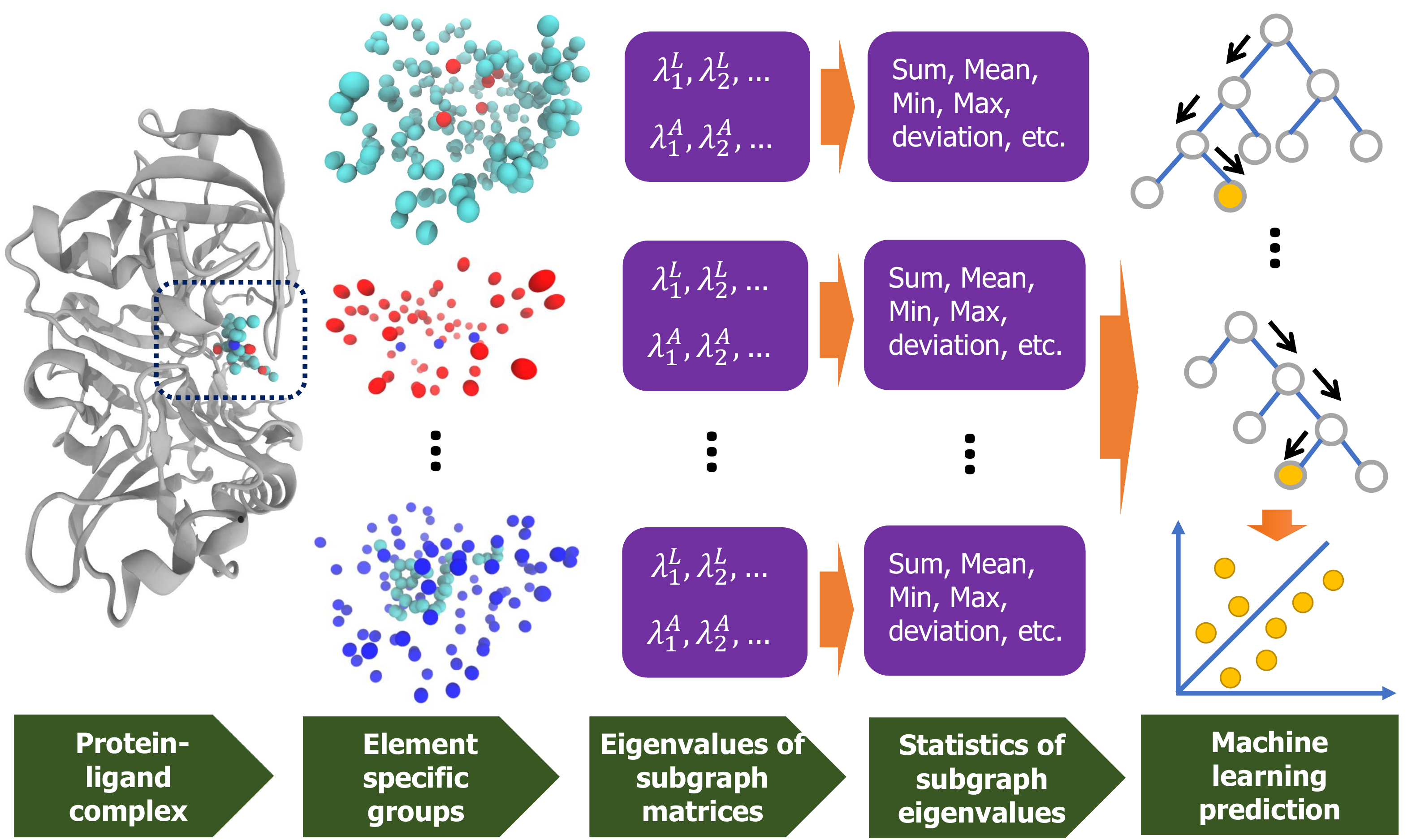}
    \caption{A paradigm of the graph-based approach. The first column is the complex input with PDBID 5QCT. The second column illustrates the element-specific groups in the binding site. The third column presents the eigenvalues of the corresponding weighted colored graph Laplacian and adjacency matrices in the second column. The statistics of these eigenvalues are calculated in the fourth column. The final column forms a gradient boosting trees model using these eigenvalues.}
    \label{fig:flowchart}
\end{figure}
The eigenvalue related information obtained from the algebraic graph approach is incorporated with machine learning algorithms to form predicting models for molecular and biomolecular properties. Depends on the nature of each learning task, regressor or classifier algorithms will be utilized. To illustrate the learning process, we denote ${\mathcal X}_i$ the $i$th structure in the training data and denote $\mathbf{G}({\mathcal X}_i; \bm{\zeta})$ a  function representing the graph information of sample ${\cal X}_i$ with respect to kernel parameters $\bm{\zeta}$. Generally, during the training process, machine learning models will minimize the following loss
\begin{equation}
\min\limits_{\bm{\zeta}, \bm{\theta}} \sum\limits_{i\in I}{\mathcal L}(\mathbf{y}_i, \mathbf{G}({\mathcal X}_i; \bm{\zeta}); \bm{\theta}),
\end{equation}
where ${\cal L}$ is the loss function, $\mathbf{y}_i$ indicates the training labels. In addition, $\bm{\theta}$ is the machine learning parameters. In principle, the set of parameters $\bm{\theta}$ will be optimized for a specific training set and the choice of a machine learning algorithm. With the current graph presentations, one can make use of advanced machine learning models such as random forest (RF), gradient boosting trees (GBTs), deep learning neural networks to minimize the loss function ${\cal L}$. To illustrate the performance of our graph-based model, we employ GBTs for a balance between accuracy and complexity. The flow chart of the proposed model is illustrated in Figure \ref{fig:flowchart}.

All the experiments in this graph learning task are carried out by the Gradient Boosting Regressor module implemented in the scikit-learn v0.19.1. The detailed parameters are given as $n$\_estimators=10000, max\_depth=7,  min\_samples\_split=3, learning\_rate=0.01, loss=ls, subsample=0.3, and max\_features=sqrt. That parameter selection is nearly optimal and is the same for all calculations.

\paragraph{Model parametrization}
Avoiding the wording, this notation  ${\rm AGL}^{\mathcal{M}}_{\Omega, \beta, \tau}$  represents the AGL-Score features encoded based on the interactive matrix type $\mathcal{M}$ along with kernel type $\Omega$  and kernel parameters $\beta$ and $\tau$.  Furthermore, $\mathcal{M}={\rm Adj}$, $\mathcal{M}={\rm Lap}$, and $\mathcal{M}={\rm Inv}$ represent adjacent matrix, Laplacian matrix, and the pseudo inverse of Laplacian matrix, respectively.  In the kernel type notation, $\Omega={\rm E}$  and $\Omega={\rm L}$, respectively, indicate generalized exponential kernel and generalized Lorentz kernels. Since the kernel order notation depends on the specific kernel type, we denote $\beta=\kappa$ if $\Omega={\rm E}$, and  $\beta=\nu$ if $\Omega={\rm L}$. Lastly, the scale factor $\tau$ i implicitly imply this expression  $\eta_{kk'}=\tau (\bar{r}_k + \bar{r}_{k'})$, in which $\bar{r}_k$ and $\bar{r}_{k'}$ are the van der Waals radii of element type  $k$ and element type $k'$, respectively.

In the multiscale representation for the AGL-Score, we naturally extend the single-scale notation. Only at most two different kernels are carrying out in the AGL-Score model, and the resulting model is denoted as  ${\rm AGL}^{\mathcal{M}_1\mathcal{M}_2}_{\Omega_1, \beta_1,\tau_1;\Omega_2,\beta_2,\tau_2}$.

\begin{table}[!ht]
    \centering
    \caption{The ranges of AGL hyperparameters for 5-fold cross-validations}
    \begin{tabular}{|c|c|}
        \hline
        Parameter & Domain\\ \hline
        $\tau$  & $\{0.5, 1.0, \dots, 6\}$ \\\hline
        $\beta$ & $\{0.5, 1.0, \dots, 6\} \cup \{10, 15, 20\}$\\\hline
        $\mathcal{M}$ & $\{{\rm Adj, Lap, Inv}\}$\\
        \hline
    \end{tabular}
    \label{tab:AGL_parameter_domains}
\end{table}

To achieve the optimal parameter selection in the AGL-Score's kernels,  we perform 5-fold cross-validation (CV) on the training data of the benchmark. Ideally, one needs to revise the machine learning model for different problem settings. To demonstrate the robustness of our graph-based features, we only train the AGL-Score's parameters on CASF-2007 benchmark with a training data size of 1105 complexes. Similar to our previous work, we select the range of the graph-based model's hyperparameters as demonstrated in Table \ref{tab:AGL_parameter_domains}. The ranges of AGL's kernel parameters are selected similarity to ones in DG-GL models discussed in Section \ref{sec:DG-GL_learning}.
For the CASF benchmark datasets, we take into account 4 atom types in protein, namely   \{${\rm C, N, O, S}$\}, and 10 atom types in the ligand, namely \{${\rm H, C, N, O, F, P, S, Cl, Br, I}$\}, that results in 40 different atom-pairwise combinations. Due to having the opposite eigenvalues in the adjacency matrix, we only consider its positive eigenvalues. Moreover, the statistical properties of these eigenvalues such as sum, minimum (i.e., the Fiedler value for Laplacian matrices or the half band gap for adjacency matrices), maximum, mean, median, standard deviation, and variance are collected.  Moreover, the number of distinct eigenvalues, as well as the summation of the second power of them, are calculated. Finally, we form a descriptor vector of 360 features.

\subsection{Machine learning algorithms}

It is generally true that our mathematical descriptors can be paired with any machine learning model. However, the devil is in the details: difference machine learning algorithms respond differently to data size, descriptor dimension, descriptor noise, descriptor correlation, descriptor amplitude, and descriptor distribution. Therefore, it is useful to design learning-model adapted mathematical descriptors.

In the past few years, we have integrated various  mathematical descriptors with a variety of machine learning algorithms, namely  k-nearest neighbors (KNNs) \cite{BaoWang:2016HPK, ZXCang:2018a}, learning to rank (LR) \cite{BaoWang:2017FFTB,BaoWang:2018FFTS}, support vector machine (SVM) \cite{ZXCang:2015},  gradient boosted decision trees (GBDT) \cite{ZXCang:2017a,ZXCang:2017b}, random forest (RF) \cite{DDNguyen:2017d, KDWu:2018a,KDWu:2018b}, extra-trees (ET) \cite{ZXCang:2018a},   deep artificial neural network (ANN) \cite{KDWu:2018a,KDWu:2018b}, deep convolutional neural network (CNN) \cite{ZXCang:2017c, ZXCang:2018a},  multitask ANN \cite{KDWu:2018a,KDWu:2018b}, multitask CNN \cite{ZXCang:2017c}, and generative networks \cite{grow2019generative}. 

Due to the extensive variability in the possible types of biological tasks and machine learning algorithms for potentially many data conditions, it is very challenging to provide an exhaustive list of fully optimized descriptors for a specific combination of biological tasks, learning algorithms and datasets. Nevertheless, one can explore near-optimal descriptors to each potential combination of biological task, learning model, and dataset and select appropriate mathematical descriptors with suitable parameters. Using topological descriptors as an example,  we outline the construction of a few topological learning strategies. In general,  kNNs are very simple and are used to facilitate optimal transport approaches, such as  Wasserstein metrics. However,  their results might not be the optimal \cite{ZXCang:2018a}. LR algorithms can be quite accurate  \cite{BaoWang:2017FFTB,BaoWang:2018FFTS}, but their training is quite time-consuming.  Ensemble methods, such as  RF, GBDT, and ET, are  relatively accurate and efficient \cite{DDNguyen:2017d, KDWu:2018a,KDWu:2018b,ZXCang:2018a}. In particular, RF should be the method of choice for a new problem due to its fewer parameters and robustness. Due to its accuracy and robustness, RF method is often used to rank the feature importance.   Utilizing a few more parameters, GBDT can typically improve RF's predictions after a more intensive parameter search.

 Ensemble methods and deep CNNs can be very accurate and robust against overfitting originated from large machine learning dimensions by shrinkage and dropout techniques, respectively \cite{ZXCang:2017a, ZXCang:2017b}. Therefore, they can be used to examine a large number of descriptors. It is worthy to note that none of these methods works well when the statistics of the test set differs much from that of the training set.   When training datasets are sufficiently large,  deep learning methods can be more accurate but might involve a very expensive training because of multiple layers of neurons \cite{KDWu:2018a, KDWu:2018b, ZXCang:2017c, ZXCang:2018a}.  Transfer learning or multitask learning can be used to improve the prediction of small datasets when they are coupled to a large dataset that shares similar statistics and the same descriptor structure \cite{KDWu:2018a, KDWu:2018b, ZXCang:2017c}.

Intrinsically low-dimensional descriptors based on advanced mathematics can be constructed for complex learning models involving multiple neural networks, such as domain adaptation, active learning, recurrent neural network, long short term memory, autoencoder,  generative adversarial networks, and various  reinforcement learning algorithms.

\section{Datasets and evaluation metrics}

\subsection{Datasets}
In this review, we illustrate our models against three commonly used drug-discovery related benchmark datasets, namely, CASF-2007 \cite{RenxiaoWang:2009Compare},  CASF-2013 \cite{YLi:2014}, and  CASF-2016 \cite{su2018comparative}. These benchmarks are collected in the PDBbind  database and have been used to evaluate the general performance of a scoring function on a diverse set of protein-ligand complexes.

\begin{table}[!ht]
    \centering
    \caption{Summary  of PDBbind datasets used in the present work}
    \begin{tabular}{|l|c|c|}
        \hline
        & Training set complexes & Test set complexes\\ \hline
        CASF-2007 benchmark   & 1105 & 195 \\\hline
        CASF-2013 benchmark   & 3516 & 195 \\\hline
        CASF-2016 benchmark   & { 3772} & { 285} \\
        \hline
    \end{tabular}
    \label{tab:PDBbind_datasets}
\end{table}

 Note that for docking power and screening power assessments, additional data information is given for
CASF-2007 \cite{RenxiaoWang:2009Compare} and  CASF-2013 \cite{YLi:2014,li2018assessing} as described in the next section.

\subsection{Evaluation metrics}
In the drug-design related benchmark, a scoring function (SF) is often validated based on four commonly metrics, namely scoring power, ranking power, docking power, and screening power \cite{RenxiaoWang:2009Compare,li2018assessing}. The following sections briefly offer introductions for these matrices and the associated datasets.

\subsubsection{Scoring power}
This metric measures how good a scoring function in predicting affinities that linearly correlate to the experimental data. To this end, the standard Pearson's correlation coefficient ($R_p$) is employed
\begin{align}
R_p = \frac{\sum\left(x_i -\bar{x}\right)\left(y_i -\bar{y}\right)}{\sqrt{\sum\left(x_i -\bar{x}\right)^2}\sqrt{\sum\left(y_i -\bar{y}\right)^2}},
\end{align}
where $x_i$ and $y_i$ are, respectively, predicted binding affinity and experimental data for the $i$th complex. The average of all predicted and experimental values are denoted as $\bar{x}$ and $\bar{y}$, respectively. All three benchmark datasets, CASF-2007, CASF-2013, and CASF-2016, were used to evaluate the scoring power of our models.

\subsubsection{Ranking power}
In this assessment, the ability to ranking the binding affinity of complexes in the same cluster is stressed \cite{RenxiaoWang:2009Compare,li2018assessing}.  Two benchmarks, CASF-2007 and CASF-2013, were used to test our AGL-Score's ranking power. Both these datasets have 65 different protein targets, and each protein has three binding distinct ligands. There two different levels of the assessments. The first is high-level success measurement which testifies if the affinities of three ligands in each cluster are correctly ranked. The other assessment is the so-called low-level success measurement which determines whether a scoring function can identify the ligand with the highest binding affinity in its cluster. The score in this assessment is calculated by the percentage of successful ranking in a given benchmark.

The above-mentioned ranking power evaluation may  not be robust since there are only three ligands in each cluster used to determine the order ranking. Thus, the real performance of the scoring function in virtual screening cannot be transferable. Moreover, more accurate statistical information can be attained by Kendall's tau or Spearman correlation coefficient as used in D3R Grand Challenges \cite{gaieb2019d3r}.

\subsubsection{Docking power}
This metric is used to testify the ability of a scoring function in discrimination the ``native'' pose from the docking software-generated structures \cite{RenxiaoWang:2009Compare}. To determine the native pose, one used the root-mean-square deviation (RMSD) between that structure and the true binding pose. If its RMSD is less than 2\AA, that pose is classified as a native. Each ligand in CASF-2007 benchmark has 100 generated structures using four docking software, namely  GOLD \cite{jones1995molecular,G-Score}, Surflex \cite{jain2003surflex,jain2007surflex}, FLexX \cite{rarey1996fast} and LigandFit \cite{venkatachalam2003ligandfit}. In CASF-2013, there are still 100 software-generated structures for each ligand but from three docking software, namely, GOLD v5.1 (https://www.ccdc.cam.ac.uk), Surflex-Dock provided in SYBYL v8.1 (https://www.certara.com/), and MOE v2011 (https://www.chemcomp.com/). It is noted that RMSD formulation in CASF-2007 is different from one in CASF-2013. Specifically, RMSD in CASF-2007 used a standard representation but property-matched RMSD (RMSD$^{\rm PM}$) is employed in CASF-2013 \cite{YLi:2014,li2018assessing}.  The use of new RMSD formulation is due to the incorrect values reported by standard RMSD on the symmetric structures. It is worthy to  mention that each ligand can have more than one ``native'' structure in the benchmark. Thus, if a scoring function can be able to detect any native poses, one can regard it as a successful task. The number of ligands whose ``native'' poses precisely selected defines the docking power of the method.

\subsubsection{Screening power}
This assessment relates to the scoring function's capability on the differentiation of a target protein's true binders from unbinding structures. CASF-2013 benchmark is used in this assessment. This dataset consists of 65 different protein classes. In each protein class, at least three ligands are binding to that target. The true binder has the highest experimental binding affinity is regarded as the best true binder. In this assessment, there are two different kinds of measurements. The first type concerns the enrichment factor (EF) in $x\%$ top-ranked candidates:
 \begin{equation}
{\rm EF}_{x\%} = \frac{\text{Number of true binders among $x\%$ top-ranked candidates}}
{\text{Total number of true binders of the given target protein}}.
\end{equation}
In this measure, top-ranked candidates are the ligands with high binding affinities predicted by the scoring function. The screen power is determined by the average of all EF values over 65 targets in the benchmark.

The second type of screening power is the success rate which concerns the best true binder identification. The percentage in identifying the best binders for 65 receptors from $x\%$ top-ranked candidates yields the value for the success rate.

\section{Results and discussions}\label{Results}

In this section, we review the scoring power, ranking power, docking power and screening power of the discussed mathematical models on the three benchmark sets including CASF-2007, CASF-2013, and CASF-2016.

\subsection{Hyperparameter optimization} \label{sec:model_hyperparams}
 
To achieve optimal hyperparameters among the possible combinations listed in Tables \ref{tab:DG-GL_parameter_domains} and \ref{tab:AGL_parameter_domains} for our models, 5-fold cross validation-based grid search strategies are taken into the account. For each CASF benchmark, the training data excluding the corresponding data is  employed for the aforementioned grid search. As a result,  the best EICs models in the differential based approach are EIC$^{HH}_{E,2,1;E,3,3}$  and EIC$^{HH}_{L,3.5,0.5;L,3.5,2}$ for CASF-2007. In CASF-2013, two optimal models are EIC$^{HH}_{E,1.5,5;E,3.5,3}$ ($R_p=0.771$) and EIC$^{HH}_{L,4.5,2.5;L,5.5,5}$. The selected hyperparameters found in CASF-2013 are also employed in CASF-2016. In AGL-Score models, we find that the following hyperparameters attain the highest cross-validations scores for all the CASF benchmarks:  ${\rm AGL}^{\rm Adj}_{{\rm E},6,2.5; {\rm E}, 4, 2}$ and ${\rm AGL}^{\rm Adj}_{{\rm L},3.5,1.5; {\rm L}, 15, 0.5}$. Noting that the consensus models, which are achieved by the mean of predictions of two associated models, will further lift the accuracy. Therefore, they are included in our experiments.

\subsection{Performance and discussion}

\subsubsection{Scoring power}

\begin{figure}[!htb]
    \centering
    \includegraphics[width=1.0\textwidth]{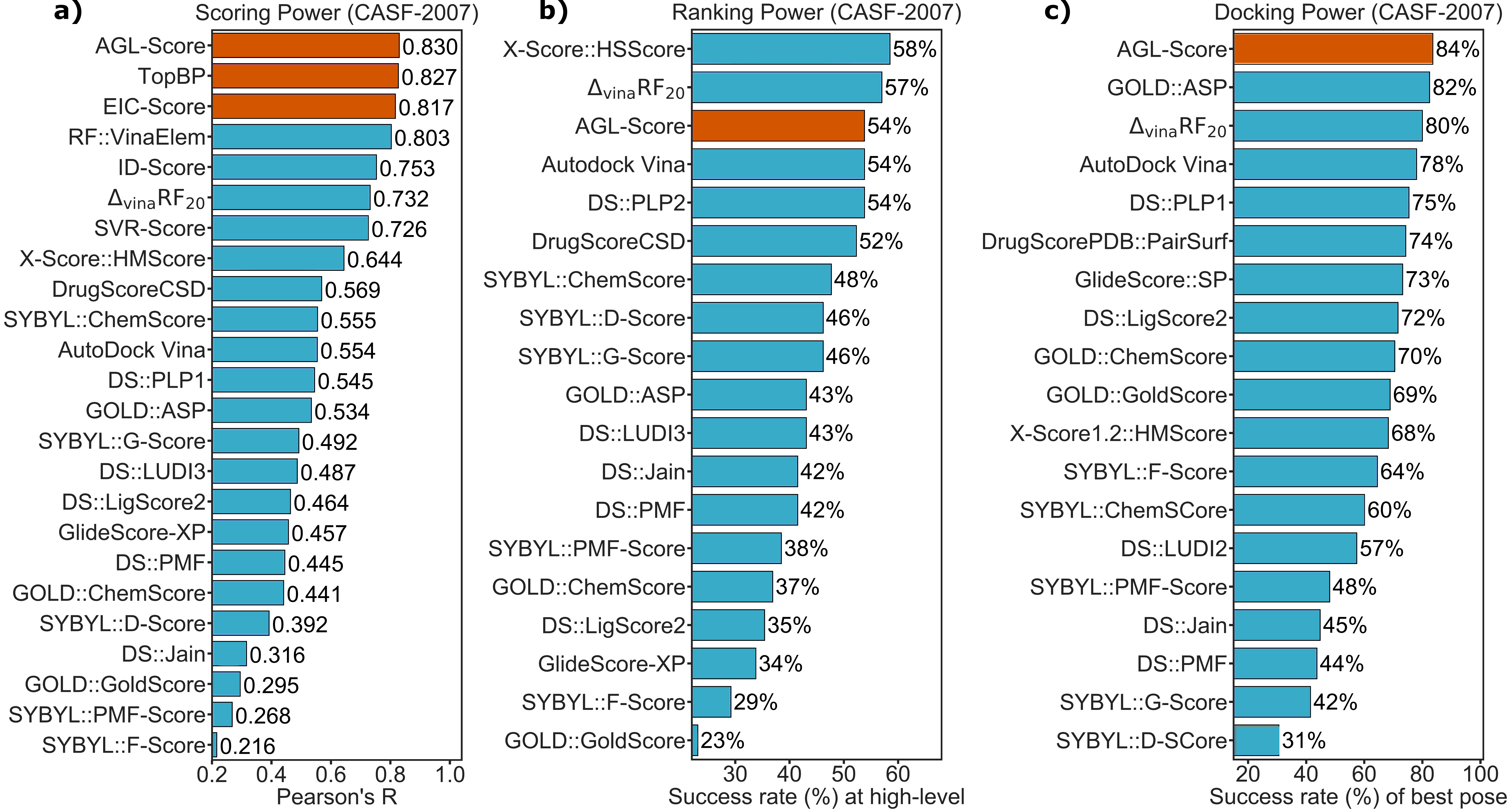}
    \caption{
The performances on different evaluation metrics of various scoring functions on CASF-2007 benchmark. a) scoring power ranked by Pearson correlation coefficient, b) ranking power assessed by the high-level success measurement, and c) docking power measured by the rate of successfully identifying the  ``native'' pose from 100 poses for each ligand. Our developed models, namely TopBP\cite{ZXCang:2018a}, EIC-Score\cite{nguyen2019dg}, and AGL-Score\cite{nguyen2019agl} are colored in orange, and other scoring functions  \cite{RenxiaoWang:2009Compare,Pedro:2010Binding, IDScore:2013,HLi:2015, wang2017improving,  nguyen2019dg, ZXCang:2017c} are colored in teal.
}
    \label{fig:v2007-benchmark}
\end{figure}


\begin{figure}[!htb]
    \centering
    \includegraphics[width=1\textwidth]{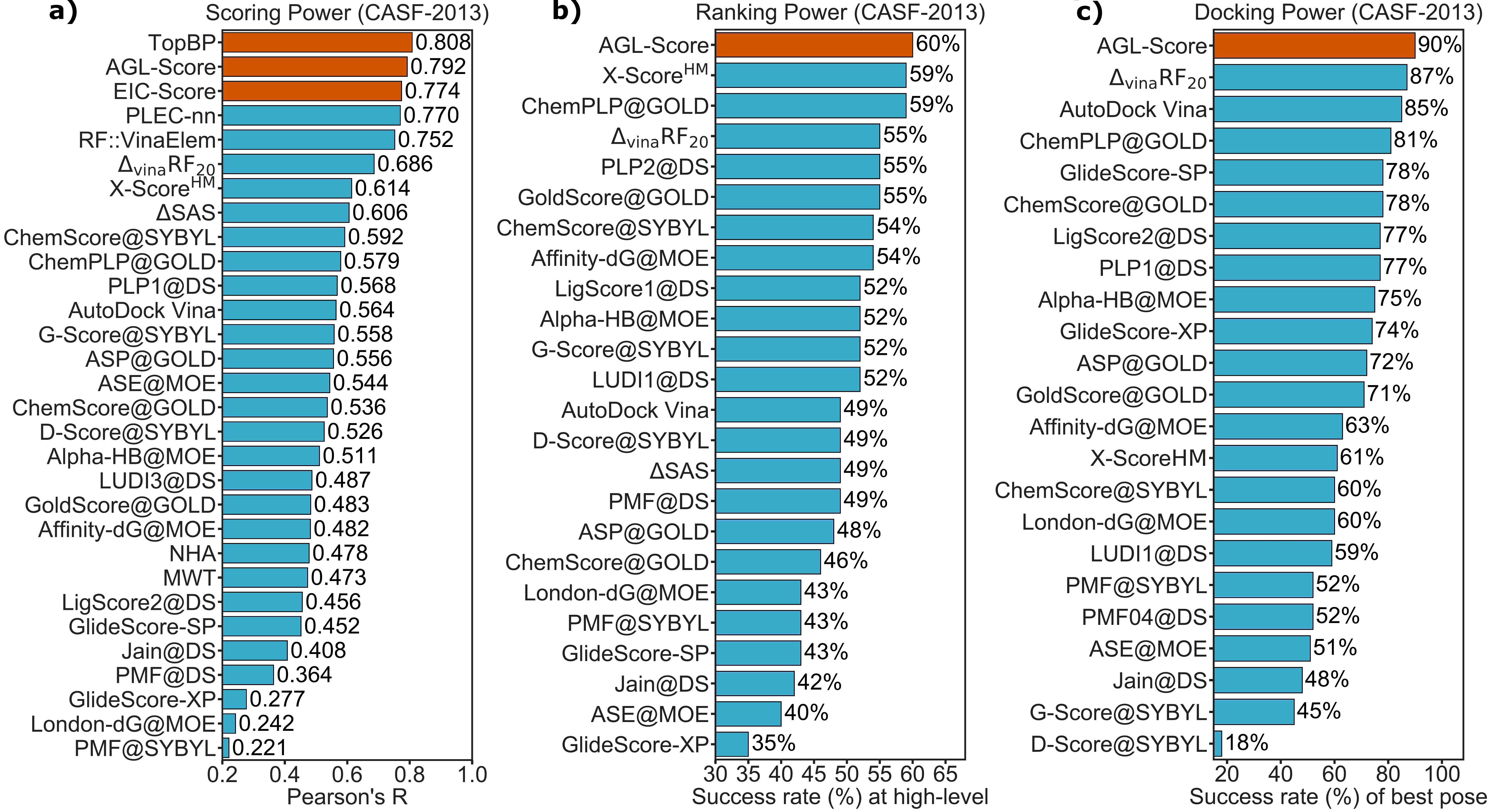}
    \caption{
    The performances on different evaluation metrics of various scoring functions on the CASF-2013 benchmark. a) scoring power ranked by Pearson correlation coefficient, b) ranking power assessed by the high-level success measurement, and c) docking power measured by the rate of successfully identifying the  ``native'' pose from 100 poses for each ligand. Our developed models, namely TopBP \cite{ZXCang:2018a}, EIC-Score \cite{nguyen2019dg}, and AGL-Score \cite{nguyen2019agl} are colored in orange, and other scoring functions   \cite{YLi:2014, HongJianLi:2015,wojcikowski2018development, wang2017improving, nguyen2019dg} are colored in teal.
}
    \label{fig:v2013-benchmark}
\end{figure}

\begin{figure}[!htb]
    \centering
    \includegraphics[width=0.9\textwidth]{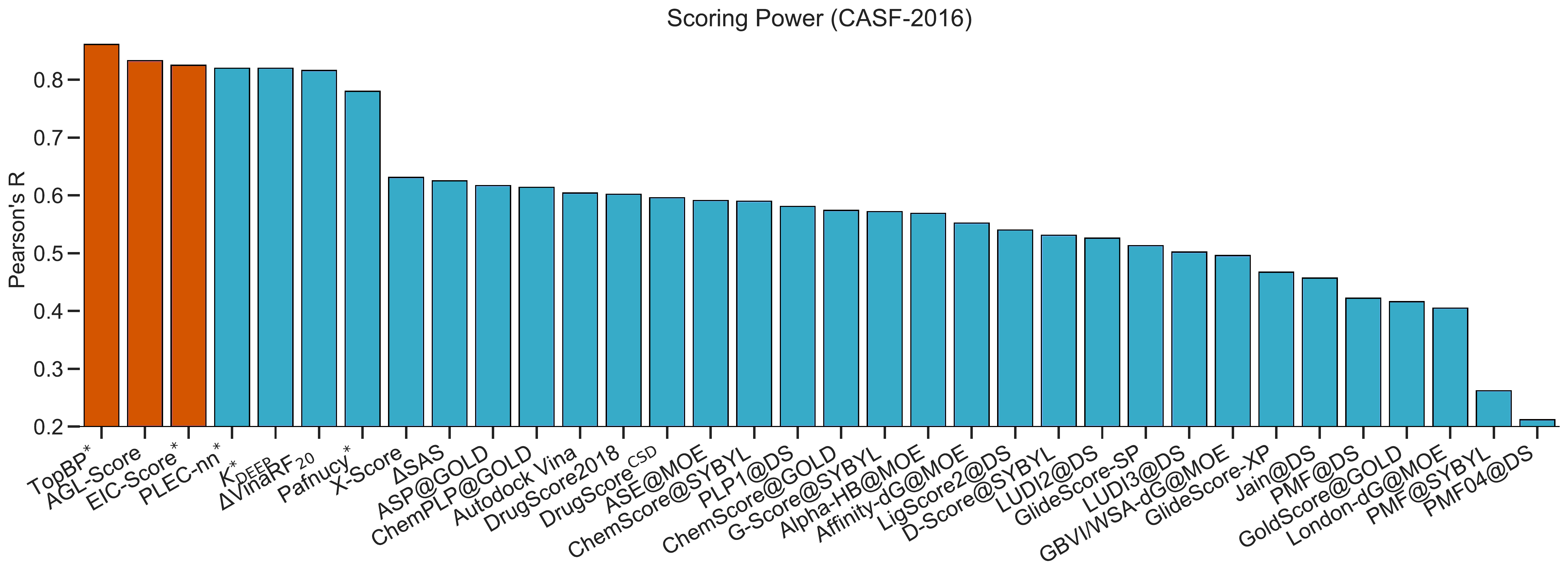}
    \caption{The Pearson correlation coefficient of various scoring functions on CASF-2016. Our developed models, namely TopBP \cite{ZXCang:2018a}, EIC-Score \cite{nguyen2019dg}, and AGL-Score \cite{nguyen2019agl} are colored
in orange. The performances of other models that are in teal are taken from Refs. \cite{su2018comparative, jimenez2018k, stepniewska2018development, wojcikowski2018development, ZXCang:2017c, nguyen2019dg}. Our TopBP is the best model with $R_p=0.861$ and ${\rm RMSE}=1.65$ kca/mol. Our AGL-Score is the second best model, with $R_p=0.833$ and ${\rm RMSE}=1.733$ kcal/mol. The third-ranked scoring function is still our model, EIC-Score, with $R_p=0.825$ and ${\rm RMSE}=1.767$ kcal/mol. Note that, scoring functions marked with $^*$ use PDBbind v2016 core set ($N=290$).}
    \label{fig:v2016-benchmark}
\end{figure}

In this task, we measure the Pearson correlation coefficient ($R_p$) between predicted affinity by our   models, namely TopBP, EIC-Score, and AGL-Score and experimental values on CASF-2007, CASF-2013, and CASF-2016. The optimal hyperparameters for AGL-Score which are chosen based on the procedure described in Section \ref{sec:model_hyperparams} are ${\rm AGL}^{\rm Adj}_{{\rm E},6,2.5; {\rm E}, 4, 2}$ and ${\rm AGL}^{\rm Adj}_{{\rm L},3.5,1.5; {\rm L}, 15, 0.5}$. To validate the scoring power of AGL-Score models on CASF-2007, we train the two aforementioned models on that benchmark's training set consisting of 1105 samples after excluding 195 complexes in the test set. To reduce the variance in our results, we perform 50 prediction task of AGL-Score models at the different random seeds. The final reported affinity is defined by averaging all the predicted values at different runs. Similarly, we also train the optimal models of DG-GL, i.e. EIC$^{HH}_{E,2,1;E,3,3}$  and EIC$^{HH}_{L,3.5,0.5;L,3.5,2}$, and Topology based models (TopBP) on 1105 complexes of CASF-2007. To compare the accuracy of our models with other state-of-the-art models, Figure \ref{fig:v2007-benchmark}a provides a comprehensive list of various scoring functions published in the literature  \cite{RenxiaoWang:2009Compare,Pedro:2010Binding, IDScore:2013,HLi:2015, wang2017improving}.  It is encouraging to see that all our models are at the top positions. Particularly, AGL-Score is the best model with $R_p=0.830$, followed by TopBP with $R_p=0.827$ and EIC-Score with $R_p=0.817$.


To predict the affinity labels of the test set consisting of 195 complexes in the CASF-2013 benchmark, we train the TopBP, EIC-Score, and AGL-Score models with optimal parameters selected in Section \ref{sec:model_hyperparams} on CASF-2013's training set having 3516 samples. We also provide a list of various scoring functions' performances on this benchmark as illustrated in Figure \ref{fig:v2013-benchmark}a. The data from that figure reveals that our TopBP is ranked in the first place with a Pearson correlation coefficient value $R_p=0.808$, followed by AGL-Score with its $R_p=0.792$. Our differential geometry-based model is in third place with $R_p=0.774$. The fourth place in the ranking table is PLEC-nn \cite{wojcikowski2018development}, a deep learning network model.

\begin{table}[!ht]
    \centering
    \caption{Discrepancy information between PDBbind v2016 core set and CASF-2016 test set}
    \begin{tabular}{|l|c|}
        \hline
        & PDBID\\ \hline
        Complexes in CASF-2016 but not in PDBbind v2016 core set   & 1g2k \\\hline
        Complexes in PDBbind v2016 core set but not in CASF-2016   & 4mrw, 4mrz, 4msn, 5c1w, 4msc, 3cyx \\\hline
    \end{tabular}
    \label{tab:PDBBind_discrepancy}
\end{table}

Similar to the training procedure on the first two benchmarks, in the last one, i.e. CASF-2016, the structures of our three models are learned from the training set ($N=3772$) of this benchmark.
Figure \ref{fig:v2016-benchmark} compares $R_p$ of numerous scoring functions on the CASF-2016. Consistently, our models still achieve the highest correlation values with $R_p=0.861$, $R_p=0.835$, and $R_p=0.825$ for TopBP, AGL-Score, and EIC-Score, respectively. It is worth noting that all top models in this benchmark are machine learning-based scoring functions, namely $K_{\rm DEEP}$ \cite{jimenez2018k}, Pafnucy \cite{stepniewska2018development}, and PLEC-nn \cite{wojcikowski2018development}. These models predict the energies for the test set of 290 complexes which is the PDBbind v2016 core set. Our topology-based model, TopBP,  was able to outperform our other methods because it used convolutional neural networks whereas  AGL-Score and EIC-Score were based on gradient boosted decision trees.

\subsubsection{Ranking power}
In this assessment, the predicted binding energies are used to determine the rank of the ligands binding to the same target. We evaluated the ranking power of three AGL-Score models, namely
generalized exponential kernel model ${\rm AGL}^{\rm Adj}_{{\rm E},6,2.5; {\rm E}, 4, 2}$and  generalized Lorentz kernel model ${\rm AGL}^{\rm Adj}_{{\rm L},3.5,1.5; {\rm L}, 15, 0.5}$, and the consensus one.  The result reveals that the generalized exponential kernel model produces the best performances on both CASF-2007 and CASF-2013 benchmarks.  Therefore, it is the representative model of the AGL-Score on this measurement. Figure \ref{fig:v2007-benchmark}b reports the ranking power of various scoring functions on CASF-2007. In this benchmark, our AGL-Score is ranked the third on high-level success with a rate of 54\%, and is behind $\Delta_{\rm vina}{\rm RF}_{20}$ (success rate = 57\%) \cite{wang2017improving} and d X-Score::HSScore (success rate = 58\%) \cite{RenxiaoWang:2009Compare}. Surprisingly, our graph-based model achieves the best success rate in CASF-2013 with the rate being 60\%, followed by X-Score${\rm HM}$ with the success rate as high as 59\%.  Since the ranking power performance depends on the predicted affinities used for the scoring power, one can see there is a correlation between these two assessments. However, our AGL-Score is the only model that is ranked in the top three places in these metrics for both CASF-2007 and CASF-2013 benchmarks.

\subsubsection{Docking power}
This docking power examines the ability of a scoring function in the discrimination between ``native'' and ``non-native'' poses. To build a robust machine learning-based model for this task, it is natural to include the diverse conformers with different range of root-mean-squared deviation (RMSD) to target experimental structure. Therefore, to create a satisfactory training data set for our AGL-Score model, we carry out GOLD v5.6.3 \cite{G-Score} to set up a training set of 1000 poses for a given  target ligand and its corresponding receptor. The parameters in the GOLD software are chosen as the following autoscale = 1.5, early\_termination = 0, and gold\_fitfunc\_path = plp. The total of computer-generated structures for both CASF-2007 and CASF-2013 benchmarks is 365,000 poses which are fed to AGL-Score for the learning process. The interested readers can download these structure information at our online server \url{https://weilab.math.msu.edu/AGL-Score}.

In considering benchmarks, each target ligand has 100 generated structures. To identify its ``native'' poses, we retrain single exponential kernel AGL-Score ${\rm AGL}^{\rm Adj}_{{\rm E},6,2.5}$ on 1000 poses generated by docking software for that specific ligand.  The single model is used here to save the calculation and training time.  The accuracy and robustness of our AGL-Score model on the docking power is illustrated in Figure \ref{fig:v2007-benchmark}c and \ref{fig:v2013-benchmark}c for CASF-2007 and CASF-2013, respectively. In both benchmarks, our graph-based model is ranked in the first place. Specifically, on CASF-2007,  the success rate of the AGL-Score model is 84\%, the second and third best models are   GOLD::ASP (82\%)\cite{RenxiaoWang:2009Compare} and $\Delta{\rm vina}{\rm RF_{20}}$ (80\%)\cite{wang2017improving}, respectively. On CASF-2013, the success rate of our method is higher with the rate being 90\%, while $\Delta{\rm vina}{\rm RF_{20}}$  \cite{wang2017improving} and Autodock Vina  \cite{wang2017improving} only reach  87\% and 85\%, respectively.

The training data of the AGL-Score model for this assessment is provided by the docking software GOLD with ChemPLP as a scoring function type (ChemPLP@GOLD).  It is interesting to see how this scoring function performs on the same benchmark. The ChemPLP@GOLD model achieves the success rates of 67\% and 82\% for CASF-2007 and CASF-2013, respectively. These values are much lower than of our model (84\% and 90\%). These comparisons confirm that our AGL-Score indeed upgrades the accuracy of the existing docking software by correctly exacting the real physical and biological properties of a biomolecular structure.

Scoring power and docking power are very two different measurement metrics. The first one concerns the affinity with the training data based on the experimental information. The latter targets the geometric validation involving artificial data. Consequently, it is not an easy task to accomplish state-of-the-art performances on both evaluations \cite{plewczynski2011can, gabel2014beware, khamis2015comparative}. According to our observation, the most commonly used docking software is reliable on identifying the ``native'' structures but inadequate in the binding energy prediction. For instance, GOLD with ASP as a scoring function (ASP@GOLD) performs quite well on the docking power with the success rate being 82\% in CASF-2007. However, ASP@GOLD's performance on the scoring power does not meet the satisfactory accuracy with $R_p=0.534$. On the contrary, the machine learning-based scoring functions often display an opposite impression. For example, RF-IChem \cite{gabel2014beware} is a machine learning model and attains a higher Pearson correlation coefficient on the scoring power ($R_p=0.791$, as expected. Unfortunately, due to the lack of proper training data and too simple descriptors for accurately encoding physical and biological information of a molecule, RF-IChem has difficulty in detecting the ``native'' pose with the success rate as low as 30\%. Recently, a machine learning-based model named $\Delta_{\rm vina}{\rm RF_{20}}$ was developed by Wang and Zhang \cite{wang2017improving} with a purpose of improving the accuracy of random-forest based scoring function on various evaluations. Indeed,  $\Delta_{\rm vina}{\rm RF_{20}}$ offers an excellent success rate (80\%) on the docking power of CASF-2007 but still shows a respectable precision on binding affinity prediction with $R_p=0.732$. Nevertheless, the Pearson correlation coefficient of the $\Delta_{\rm vina}{\rm RF_{20}}$  is far behind the elite models such as  TNet-BP ($R_p$ = 0.826) \cite{ZXCang:2017c}. Our graph based-model , AGL-Score, not only has a great accomplishment on the docking power (success rate = 84\% in CASF-2007) as   $\Delta_{\rm vina}{\rm RF_{20}}$ , but also performs similarly to  TNet-BP on the scoring power ($R_p=0.83$ in CASF-2007). These results again reinforce the ability of the AGL-Score in capturing the crucial interactions in molecular and biomolecular structures.
\begin{figure}[!htb]
    \centering
    \includegraphics[width=0.8\textwidth]{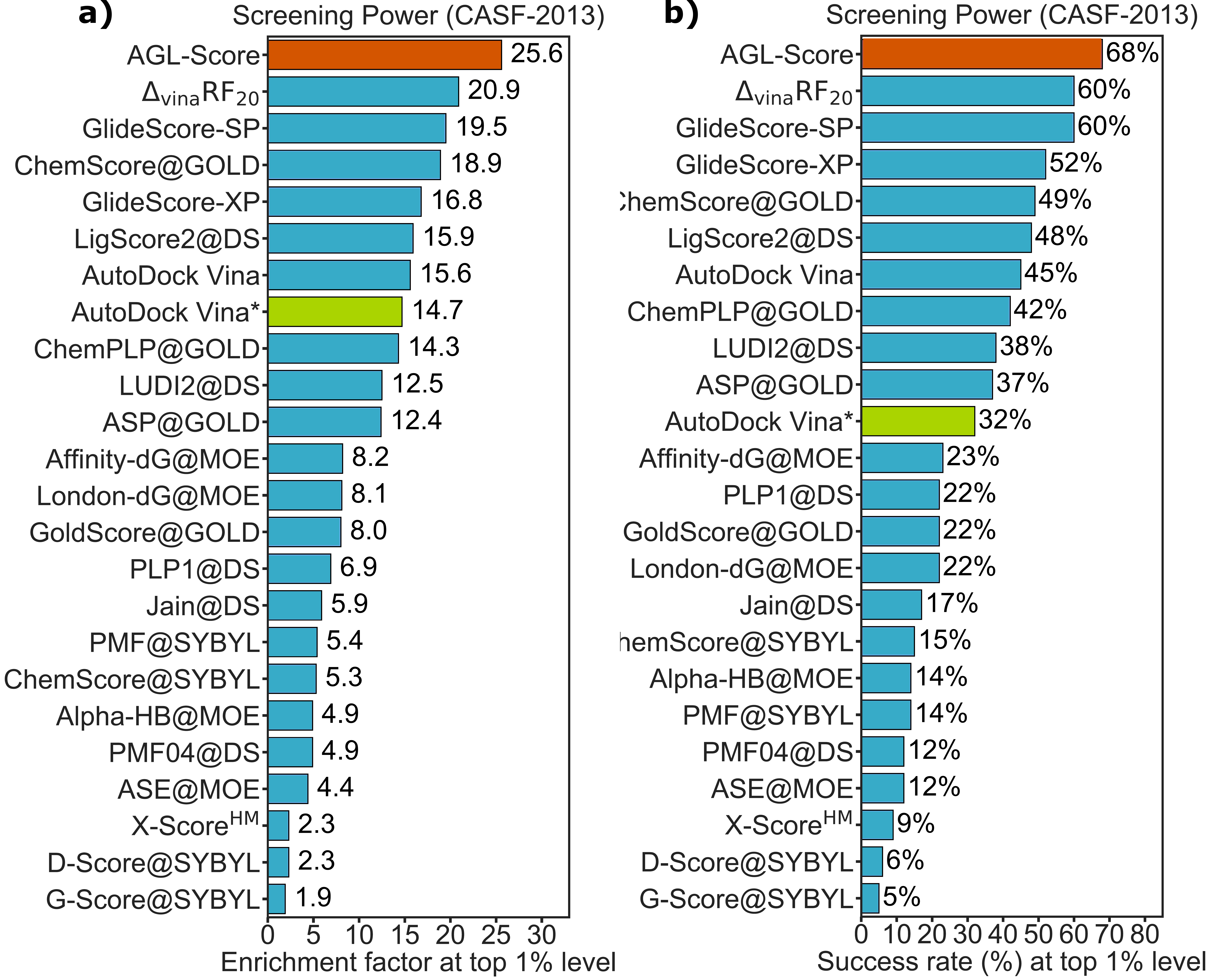}
    \caption{{
The performances of various scoring functions on the screening power for CASF-2013 benchmark based on a) enrichment factor and b) success rate at the top 1\% level. The orange bar indicates our graph-based models \cite{nguyen2019agl}. The green bar represents the results of Autodock Vina carried out in our lab. The teal bars express the performances of other models Refs. \cite{YLi:2014, wang2017improving}.
}}
    \label{fig:v2013-SP}
\end{figure}

\subsubsection{Screening power}

In this assessment, we verify the ability of the AGL-Score in picking up the true binders for different 65 protein classes in the CASF-2013 benchmark. The power metric concerns the active and inactive of 195 ligands for a specific class of protein rather than the estimation of a binding affinity for an experimental complex or ``native'' conformer identification. Therefore, to effectively carry out the machine learning scoring function on this take, one needs to construct an appropriate training data tailoring the active/inactive classification purpose. To this end, our training data consists of docking software-generated poses and corresponding energies.  The 3D structures of 195 ligands binding to a specific target are also created by the docking program and their energies are estimated by our AGL-Score model. The predicted true binders are identified based on their predicted affinities.

Our training set for AGL-Score on this screen power test is based on the PDBbind v2015 refine set excluding the core set in that database. Besides these experimental structures, we generate the non-binder structures for each target protein by using Autodock Vina \cite{Trott:2010AutoDock}. Specifically, we use that docking software to dock all ligands in the PDBbind v2015 refined set without the inclusion of the core-set compounds to the interested receptor. Here are the parameters of Autodock Vina we use in this procedure: exhaustiveness=10, num\_modes=10, and energy\_range=3. For each docking run, the pose associated with the highest predicted affinity by Autodock Vina is kept.

To preserve the consistency in the energy unit, all the Autodock Vina scores in kcal/mol are converted to pKd unit via a constant factor -1.3633 \cite{BaoWang:2017FFTB}. Ligands in the PDBbind v2015 refined set which do not bind to a target protein are designated as decoys  \cite{YLi:2014,li2018assessing}. To conserve the physical and biological sense, the Autodock Vina predicted energies of those decoys cannot be higher than the lowest energies among the ligands experimentally bind to that target protein.  To this end, we constraint the decoy energies by the lower bound of the true binders. The generated structures, as well as the energy labels of the decoys used in the AGL-Score training process, are publicly available at \url{https://weilab.math.msu.edu/AGL-Score}.

The AGL-Score model we used in this screening power is AGL-Score ${\rm AGL}^{\rm Adj}_{{\rm E},6,2.5} $. Figure \ref{fig:v2013-SP} plots the performance of the AGL-Score along with numerous scoring functions reported in the literature \cite{YLi:2014, wang2017improving}. It is an encouragement to see our AGL-Score achieves the top performance on enrichment factor (EF) and success rate at the top 1\% level in the CASF-2013 benchmark. The EF of the AGL-Score is 25.6 followed by  $\Delta_{\rm vina}{\rm RF_{20}}$ (EF=20.9)\cite{wang2017improving} and GlideScore-SP (EF=19.5)\cite{YLi:2014}. Moreover, the success rate of our graph-based model is 68\% followed by $\Delta_{\rm vina}{\rm RF_{20}}$ and GlideScore-SP that both attain 60\%.

Since the partial training data of our AGL-Score model is generated by Autodock Vina, it is interesting to see the accuracy of that docking software carried out in our lab on this assessment. The Autodock Vina's performances are much lower than the graph-based model. Specifically, the docking software attains EF as low as 14.7 while AGL-Score produces EF as high as 25.6.  In the success rate metric, Autodock Vina's accuracy is only 32\% which is far from AGL-Score's rate at 68\%. Since the published work \cite{wang2017improving} already reported Vina's screen power tests, to avoid any confusion we plot our experiments on the Vina software as green bars in Figure \ref{fig:v2013-SP}.  The unsatisfactory results of the Autodock Vina on the screen power further reinforce the accurately encoded physical and biological information in our graph-based model rather than the dependence on training quality.

 The screening power validation is an important metric in virtual screening in drug design. Since this assessment strictly requires meaningful molecular descriptors and an appropriate training set, large numbers of machine learning-based scoring functions with simple features and irrelevant training data often perform poorly on this metric despite the promising accuracy on the scoring power. For instance, RF@ML \cite{khamis2015comparative} is a machine learning model using Random Forest for the prediction but its features simply count the number of intermolecular contacts between two atom types. In fact, RF@ML produces an acceptable correlation ($R_p=$0.704) on 164 complexes in PDBbind v2013 dataset . However, RF@ML's accuracies of screen power are the worst among the models listed in Figure \ref{fig:v2013-SP}. In contrast, our AGL-Score model with superior feature representations and training data insight has achieved the top places in both scoring and screening powers.

\subsection{Online servers }\label{Servers}
In the past few years, a few online servers have been developed for  the predictions of protein-ligand binding affinities
(\href{https://weilab.math.msu.edu/RI-Score/}{RI-Score},
\href{https://weilab.math.msu.edu/TML/TML-BP/}{TML-BP},  and
\href{https://weilab.math.msu.edu/TML/TML-BP/}{TML-BP}),
protein stability changes upon mutation
(\href{https://weilab.math.msu.edu/TML/TML-MP/}{TML-MP}, and
\href{https://weilab.math.msu.edu/TML/TML-MP/}{TML-MP}),
molecular toxicity
(\href{https://weilab.math.msu.edu/TopTox/}{TopTox}),
partition coefficient and aqueous solubility
(\href{https://weilab.math.msu.edu/TopP-S/}{TopP-S}),  and
protein flexibility (\href{https://weilab.math.msu.edu/FRI/}{FRI}).


\section{Concluding  remarks}\label{Conclusion}

Artificial Intelligence (AI), including machine learning (ML), has had tremendous impacts on science, engineering, technology,  healthcare, security, finance, education, and industry, to name just a few. However, the development of ML algorithms for macromolecular systems is hindered by their intricate structural complexity and associated high ML dimensionality. In the past few years, we have addressed these challenges by three classes of mathematical techniques based on algebraic topology, differential geometry, and graph theory. These mathematical apparatuses are enormously effective for macromolecular structural simplification and ML dimensionality reduction. By integrating with advanced ML algorithms,  we have demonstrated that our mathematical approaches give rise to the best prediction in D3R Grand Challenges, a worldwide competition series in computer-aided drug design \cite{nguyen2019mathematical,nguyen2019mathdl}, as well as many other physical, chemical and biological datasets.
Nonetheless, our methods and results were scattered over a number of papers. In this review, we provide a systematical and coherent narration of our state-of-the-art algebraic topology, differential geometry, and graph theory-based methods. Emphasis is given to the physical and biological challenge-guided evolution of these mathematical approaches. Although our mathematical methods can be paired with various machine learning algorithms for a wide variety of chemical, physical, and biological systems, we focus on protein-ligand binding analysis and prediction in the present review.

Fueled by the fast advances in  ML and the availability of biological datasets,  recent years witness the rapid growth in the development of advanced mathematical tools in the realm of molecular biology and biophysics.  In most of history, mathematics has been the driving force for natural science. Indeed, mathematics is the underpinning for every aspect of modern physics, from electrodynamics, thermodynamics, statistical mechanics, quantum mechanics,  solid state physics, quantum field theory, to the general theory of relativity. In the past century, mathematics and physics have been mutually beneficial.  Similar, mathematics will become an indispensable part of biological sciences shortly. Currently, algebraic topology, differential geometry, graph theory, group theory,  differential equations, algebra,  and combinatorics have been widely applied to biological science. Many other advanced mathematical subjects, such as algebraic geometry and low dimensional manifolds will soon find their applications to biological science.

The next generation of AI and ML technologies will be designed to understand the rules of life and reveal the physical and molecular mechanics of biomolecular systems.  Such a development will bring tremendous benefits to health sciences, including drug discovery. Mathematics will play a paramount role in future AI and ML technologies. On the one hand, the mathematical theory will contribute to the foundation of AL and the design principle of ML. On the other hand,  new mathematical descriptors will be developed to enable the automatic discovery of scientific laws and principles \cite{schmidt2009distilling}.  New mathematical descriptors will be made physically interpretable so that machine learning predictions from these descriptors can reveal new molecular mechanisms. A generation of new mathematical descriptors will be made adaptive to future AI technology.   Mathematical descriptors will be systematically validated and optimized on a vast variety of existing datasets.

 \vspace{1cm}

 \section*{Acknowledgments}
This work was supported in part by  NSF Grants DMS-1721024, DMS-1761320, and IIS1900473,   NIH grants  GM126189 and  GM129004,  Bristol-Myers Squibb, and Pfizer. We thank Dr. Kaifu Gao for his contribution to our team's pose prediction in D3R Grand Challenge 4. 

\vspace{1cm}

\section*{Literature cited}
\renewcommand\refname{}


\end{document}